\pdfoutput=1
\documentclass[twocolumn,longbibliography,superscriptaddress,floatfix]{revtex4-1}

\usepackage{graphicx}% Include figure files
\usepackage{dcolumn}% Align table columns on decimal point
\usepackage{mathtools, nccmath}
\usepackage{bm}
\usepackage{xcolor}
\usepackage{amsmath}
\usepackage[export]{adjustbox}
\usepackage{framed}
\usepackage[]{graphicx}
\usepackage[caption=false]{subfig}
\usepackage{hyperref}% add hypertext capabilities
\usepackage[mathlines]{lineno}% Enable numbering of text and display math
\usepackage{textcomp}
\usepackage{gensymb}
\usepackage[super]{nth}
\begin{document}

\preprint{APS/123-QED}

\title{Pseudo field effects in type II Weyl semimetals: new probes for over tilted cones}
\author{Daniel Sabsovich}
\affiliation{Raymond and Beverly Sackler School of  Physics and Astronomy, Tel Aviv University, Tel Aviv 69978, Israel}
\author{Tobias Meng}
\affiliation{Institute of Theoretical Physics, Technische Universitat Dresden, 01062 Dresden, Germany}
\author{Dmitry I. Pikulin}
\affiliation{Microsoft Quantum, Station Q, University of California, Santa Barbara, CA 93106, USA}
\affiliation{Microsoft Quantum, Redmond, Washington 98052, USA}
\author{Raquel Queiroz}
\affiliation{Department of Condensed Matter Physics, Weizmann Institute of Science, Rehovot 7610001, Israel}
\author{Roni Ilan}
\affiliation{Raymond and Beverly Sackler School of  Physics and Astronomy, Tel Aviv University, Tel Aviv 69978, Israel}
\date{\today}% It is always \today, today,
             %  but any date may be explicitly specified

\begin{abstract}
We study the effects of pseudo-magnetic fields on Weyl semimetals with over-tilted Weyl cones, or type II cones. We compare the phenomenology of the resulting pseudo-Landau levels in the type II Weyl semimetal to the known case of type I cones. We predict that due to the nature of the chiral Landau level resulting from a magnetic field, a pseudo-magnetic field, or their combination, the optical conductivity can be utilized to detect a type II phase and deduce the direction of the tilt. Finally, we discuss ways to engineer homogeneous and inhomogeneous type II semimetals via generalizations of known layered constructions in order to create controlled pseudo-magnetic fields and over-tilted cones. 
\end{abstract}

\maketitle
%\tableofcontents

\section{Introduction}

The recent advent of Dirac materials has forged new connections between high energy physics and condensed matter \cite{Volovik2003}. Dirac materials are characterized by low-energy Hamiltonians that take the form of a Dirac equation \cite{Dirac1928}. Prototypical examples include graphene \cite{Castro2009}, and the surface states of topological insulators \cite{Bernevig2013,Hasan2010,Qi2011,Kane2011,Moore2010}, both of which are governed by two-dimensional Dirac Hamiltonians. Similar physics also occurs in three dimensions, where a plethora of compounds exhibiting low-energy quasiparticles governed by variants of the Dirac or Weyl equations \cite{Weyl1929,Burkov2011b,Xiangang2011,turner2013,Hosur2013,Armitage2018}. This advancement in material science provides an avenue for testing unobserved effects attributed to high-energy particles \cite{Peskin1995,Grushin2012}. Weyl fermions, to date only confirmed as quasiparticles in solids, were theorized to exhibit a chiral magnetic effect and a chiral anomaly \cite{Nielsen1983,Fukushima2008,Zyuzin2012,Vazifeh2013}. Both of these phenomena have recently been observed in such solids \cite{Armitage2018,Binghai2017,Huang2015Chiral}. In addition, signatures of the mixed-axial gravitational anomaly were reported in thermal transport \cite{Gooth2017,Gooth2018}. The translation of high-energy results into a solid state language holds the promise of enabling thus far unrecognized functionalities that can be unlocked by engineering solid state devices.

Phenomena emerging in Dirac materials are, however, more than a realization of high-energy analogues. The equations governing the dynamics of particles in solids are in many ways more complex and less restrained than the equations associated with those of the true high-energy particles. One prominent example is a tilted Dirac or Weyl node: the tilting of the node translates to a broken Lorentz symmetry and has no analogue in high-energy physics \cite{Yong2015,Soluyanov2015,Volovik2014}. Solids therefore allow to explore fundamentally novel effects. This work deals with an important prerequisite for exploiting tilts in solid state devices, namely their detection in transport measurements. We propose to use the interplay of magnetic fields and pseudo-magnetic fields \cite{Vozmediano2015,Grushin2016,Pikulin2016,Ilan2019} in Weyl semimetals categorized as type II to fingerprint tilts in transport measurements.

A key player in our study are intrinsic pseudo-magnetic fields that emerge in Weyl materials. Such pseudo (or axial) fields mimic axial electromagnetic fields in the vicinity of Weyl points, but couple with opposite signs to nodes of opposite chiralities. In a solid, pseudo-fields can be created by a spatially or temporally varying Weyl node separation \cite{Ilan2019}. Like regular electromagnetic fields, pseudo-fields can drive variants of the chiral anomaly and the chiral magnetic effect \cite{Zhou2013,QiXiaoLiang2013,Grushin2016,Pikulin2016,Cortijo2016,Landsteiner2016,Huang2017,Ilan2019,Behrends2019}. As we show here, the combination of external magnetic fields, pseudo-magnetic fields, and a tilt of the Weyl cones affects transport in specific ways allowing to identify the tilt of the cones. More precisely, we specify measurement protocols that allow to differentiate between tilted and over-tilted cones, and pinpoint the extent and direction of the tilt. 
 
The basis for our observation is the ability to flip the chirality of the lowest Landau levels in over-tilted cones, as was pointed out before in Refs.~\cite{Soluyanov2015,Udagawa2016}: Landau levels of Weyl fermions feature a single, linearly dispersing chiral level in the vicinity of the nodes. The group velocity of all states within that Landau level is determined by the direction of the field, and the chiral topological charge of the nodes. As we show here, a combination of external and intrinsic fields can be used to control such chirality flips in a way that leaves hallmarks on the optical conductivity, allowing to resolve the arrangement of the cones.

Despite the important differences between Weyl semimetals of type I and type II laid out above, extracting evidence for the over-tilt of cones from transport experiments has been proven challenging \cite{Ma2019,Hu2019}. This is because the main transport probe of Weyl semimetals, proposed so far, is the magneto-resistance, arguably enhanced by the chiral anomaly \cite{Son2013,Knoll2019}. The magneto-resistance is, however, somewhat unspecific in regard to tilts as it shows anisotropy and large enhancement with field for both semimetals of type I and type II, and even in materials without well-defined chiralities \cite{Reis2016,Liang2018,Hu2019}. In the following, we discuss an alternative magnetotransport protocol that facilitates the reliable detection of the tilted Weyl nodes. This protocol could in principle be implemented in the magnetic type II Weyl semimetal YbMnBi$_2$ \cite{Borisenko2019}. This material features two tilted nodes, ideal for testing the effects we describe below. Nevertheless, it is not yet clear how to engineer the necessary gradient of the node separation without affecting other degrees of freedom in this complex material. We therefore also propose two types of multilayer hetero-structures in the spirit of those previously proposed by Burkov, Balents, and others \cite{Burkov2011,Meng2012,Lau2017}. These heterostructures enable the targeted design of an over-tilted inhomogeneous semimetal with two or four nodes, breaking or respecting time reversal symmetry. The first of the structures is based on magnetic hetero-structures combined with strong topological insulators, and the second is based on stacking topological crystalline insulators. The ability to engineer inhomogeneities in Weyl semimetals is also important beyond the scope of the present work: spatially varying tilts have recently been linked to the physics of black hole analogues  \cite{Volovik2016,Guan2017}. The layer constructions we present here can be extended to include such effects, and we leave the exploration of those to future work.

In addition to differentiating type I and II semimetals, our results also have experimental  implications for the detection of pseudo fields. As a consequence of the relation between our prediction for the optical conductivity and the behaviour of the lowest Landau levels generated by pseudo-fields, the measurement protocols we suggest can also lead to an experimental signature of pseudo-fields in 3D condensed matter systems, which are currently lacking.

\section{Lowest Landau level chirality flips: external magnetic fields vs. axial magnetic fields}

At zero magnetic field $\mathbf{B}=0$, the minimal low-energy model describing the dynamics of quasiparticles around a Weyl node is given by the Hamiltonian
\begin{equation}
H_{\mathbf{B}=0}=k_{i}\,v_{ij}\,\sigma_{j}+g_{i}\,k_{i}\,\sigma_{0},
\label{eq:Weyl_H}
\end{equation}
where $i,j\in{x,y,z}$, $\sigma_i$ are Pauli matrices, $\sigma_0$ is the identity matrix, and $v_{ij}$ denotes the velocity matrix. We from now on specialize to the case $v_{ij} = v_i\,\delta_{ij}$. The tilt of the Weyl node is determined by $g_{i}$: for vanishing $g_i$, the above Hamiltonian takes the form of the standard Weyl equation, whereas finite $g_i$ implies a tilting of the nodes. For $g_i>v_i$, the Weyl semimetal is said to be of type II: the cones are over-tilted and a finite Fermi surface exists even at the Weyl point \cite{Yong2015,Soluyanov2015,Volovik2014}. Otherwise, the semimetal is of type I, with a vanishing density of states at the Weyl point.

We begin by introducing the Landau levels created when a Weyl semimetal experiences a magnetic field, their dependence on the chiral charge of a Weyl node, and the effect of a tilting of the cones, including chirality flips. To make the discussion less cumbersome, we first consider a Weyl semimetal hosting only a single pair of Weyl nodes with opposite chiralities and a magnetic field which is aligned in the direction of the tilt. In addition, we start from a discussion of two \textit{co-tilted} nodes, namely node that tilt in the same direction. Later on we generalize our discussion to include counter-tilted cones and a larger number of nodes. We start out by discussing the low energy theory and then support all observations by exploring a simple lattice model.  

\begin{figure}[t]
\captionsetup[subfigure]{labelformat=empty}
\captionsetup[subfloat]{farskip=0pt,captionskip=1pt}
\centering
\subfloat[]{\includegraphics[width=0.5\columnwidth]{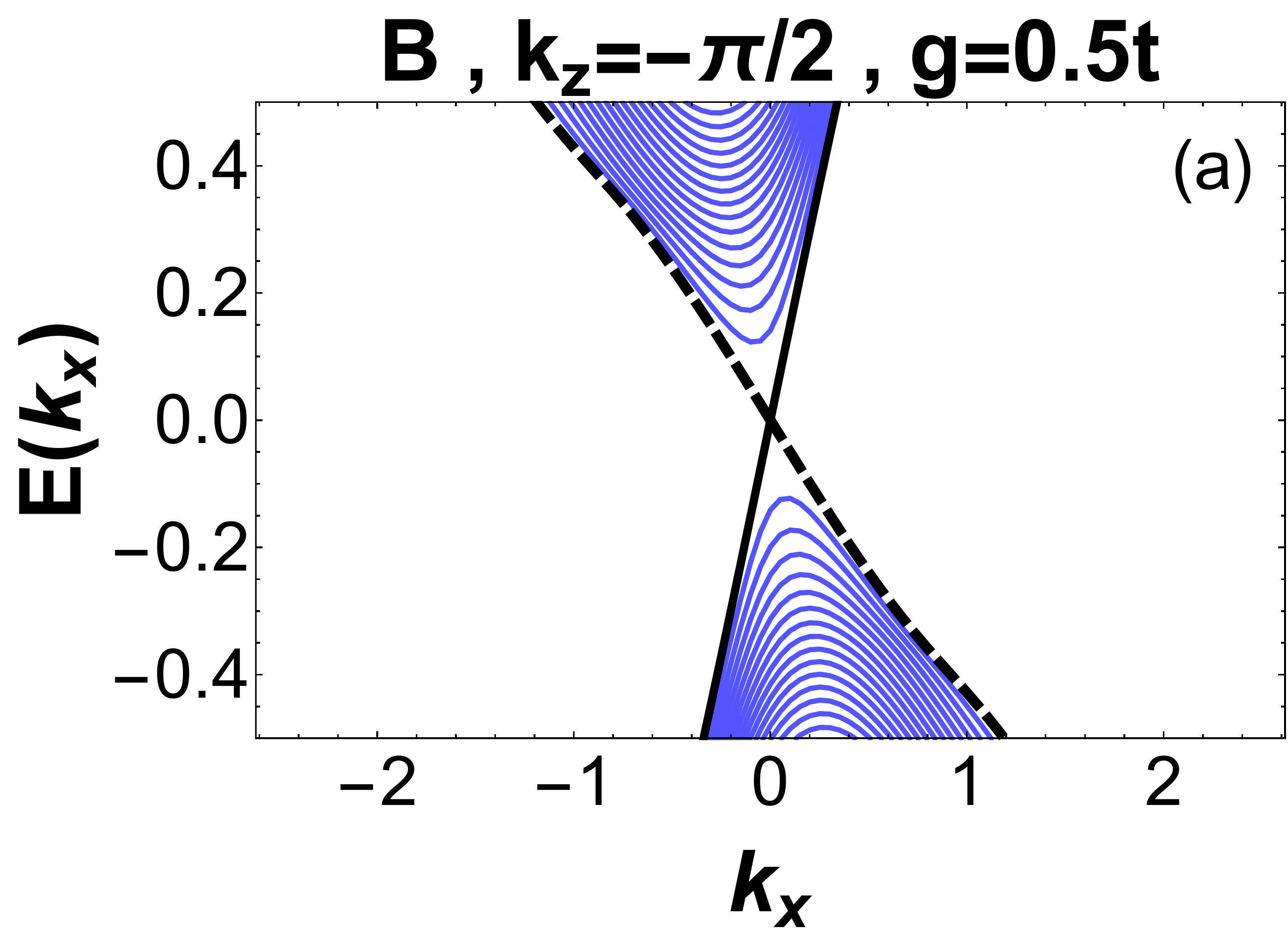}}\hfill
\subfloat[]{\includegraphics[width=0.5\columnwidth]{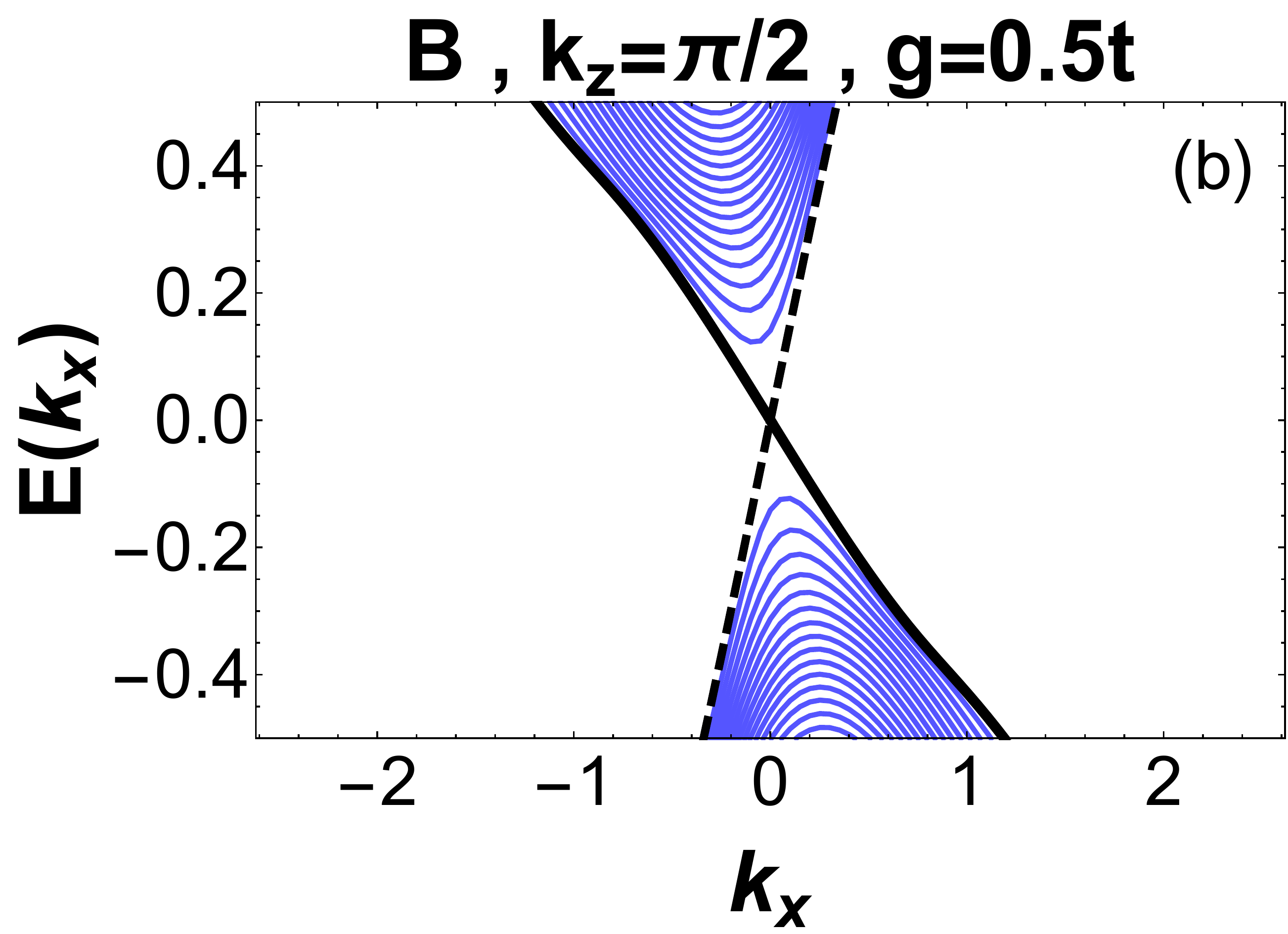}}\hfill
\subfloat[]{\includegraphics[width=0.5\columnwidth]{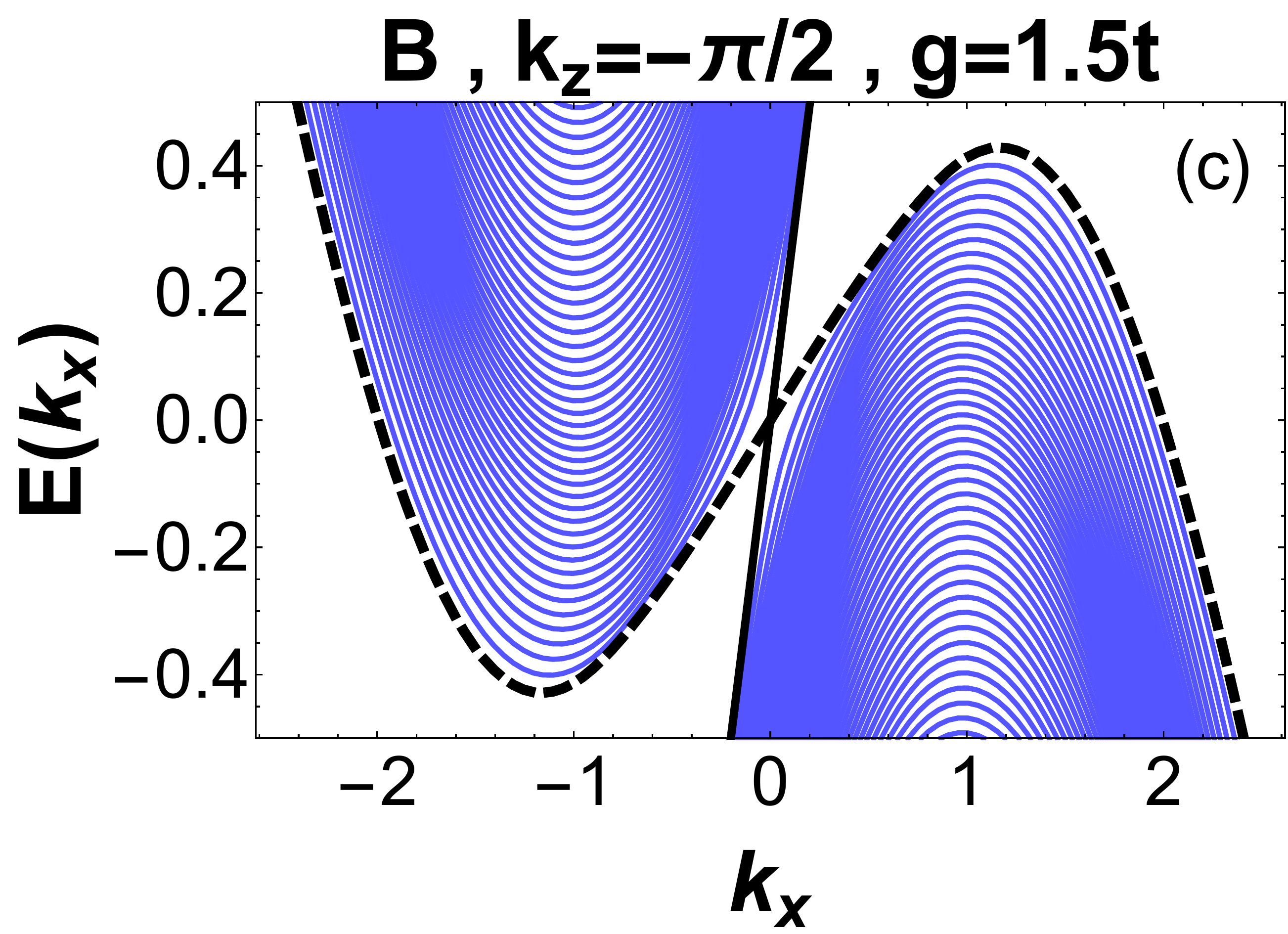}}\hfill
\subfloat[]{\includegraphics[width=0.5\columnwidth]{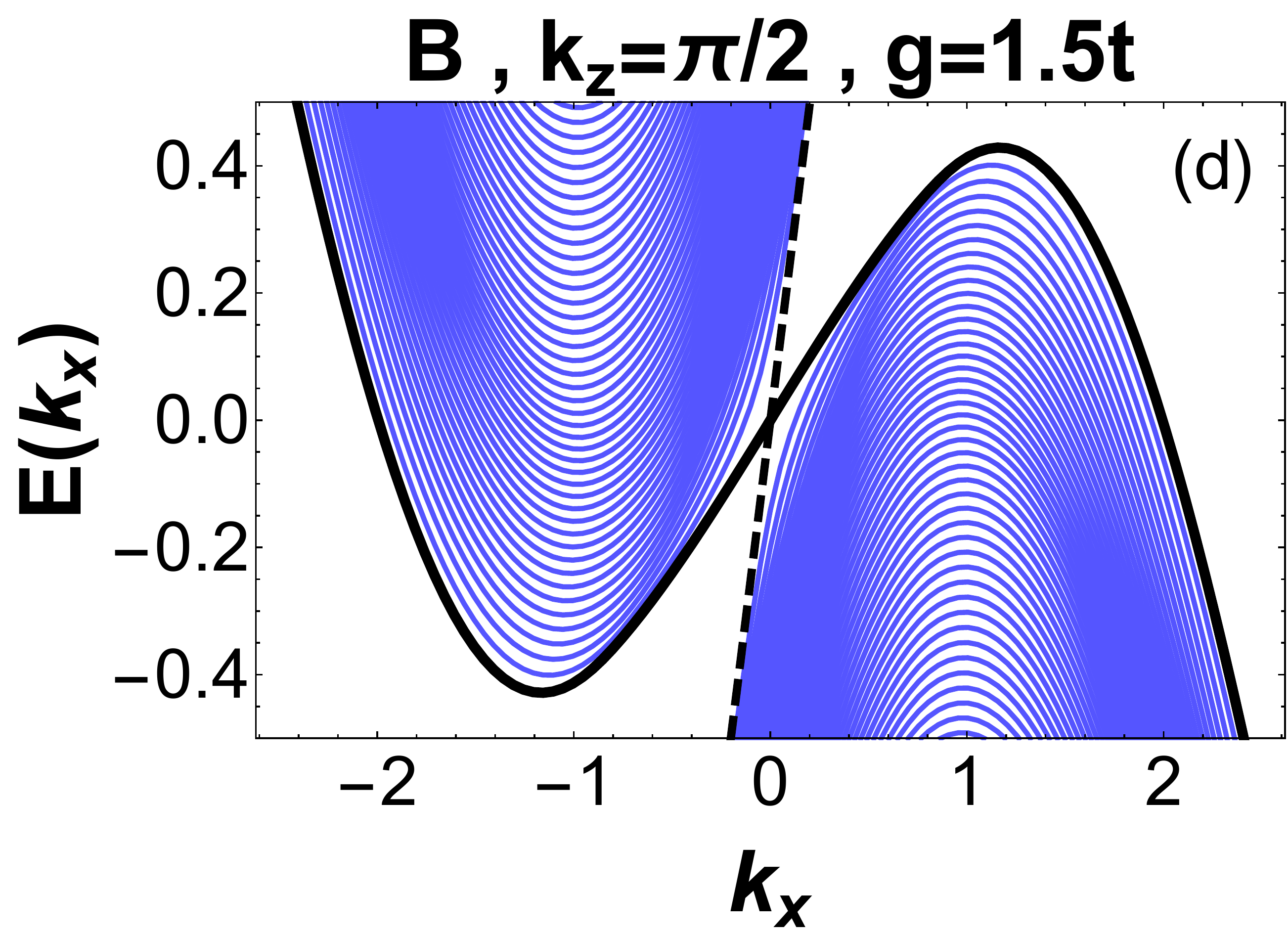}}\hfill
\vspace{-0.2in}\caption{$E(k_{x})$ Spectrum of a Weyl semimetal given by Eq.~$\ref{eq:TBModel}$ in the vicinity of the Weyl points and in the presence of a constant external magnetic field. Here, $B=0.01\hat{x}$, with $a=1$, $M=2$, $n=101$ for different values of $g$. $n\ne0$ bulk modes are colored $\boldsymbol{blue}$, $n=0$ bulk modes are in $\boldsymbol{solid\:black}$ and surface modes in $\boldsymbol{dashed\:black}$. For (a),(b) $g=0.5t$ (type I) and the spectra are evaluated close to $k_{z}=-\pi/2,=\pi/2$ respectively. for (c),(d), $g=1.5t$ (type II) and spectrum is evaluated around $k_{z}=-\pi/2,\pi/2$ respectively. One Bulk LL experiences a chirality flip close to the node. }\label{fig:WeylEkxB}
\end{figure}

\subsection{Low-energy theory}
An external magnetic field enters the Hamiltonian via minimal coupling, such that the low-energy Hamiltonian in Eq. \ref{eq:Weyl_H} becomes
\begin{eqnarray}
H_{\mathbf{B}}=C v_F\boldsymbol{\sigma}\cdot\left(-i\nabla+e\boldsymbol{A}(\mathbf{r})\right)+g\,(-i\partial_{\parallel})\,\sigma_0 ,
\label{eq:WeylHamiltonianB}
\end{eqnarray}
where $C=\pm 1$ is the chirality of the node, and where we assumed an isotropic Fermi velocity $|v_{ij}| = \delta_{ij}v_F$. $\mathbf{A}$ is the vector potential, and $\partial_\parallel$ denotes the partial derivative in the direction of the tilt and the magnetic field. The field then splits the spectrum into Landau levels with energies \cite{Udagawa2016,Yu2016}
\begin{eqnarray}
E_{n,\pm}\left(k_{\parallel}\right)=\pm v_F\sqrt{k_{\parallel}^{2}+2\left|\boldsymbol{B}\right|n}+g k_{\parallel} ,
\label{eq:WeylSpectrumnB}
\end{eqnarray}
for $n\ge1$, and
\begin{eqnarray}
E_{0,C}\left(k_{\parallel}\right)=(g-C v_F)k_{\parallel} , \label{eq:WeylSpectrum0B}
\end{eqnarray}
for $n=0$. Here,  $k_{\parallel}$ denotes the momentum in the direction of the magnetic field. Because the Landau levels only disperse with the momentum parallel to the tilt and the magnetic field, they can be understood as effectively one-dimensional modes. The zeroth Landau levels are special since they have a unique direction of propagation defined by their group velocity, while all other Landau levels contain modes of both positive and negative group velocities. In the absence of a tilt, $g=0$, the chirality of the zeroth Landau levels is opposite for nodes of opposite chiralities, as follows from Eq. \ref{eq:WeylSpectrum0B}. 

This one-dimensional chirality is robust to small variations of the Hamiltonian. In particular, this also includes small tilts $|g|<v_F$, which merely lead to a different magnitude of the group velocities at the two nodes. Stronger tilts, however, can lead to qualitatively new physics. In particular, when the Weyl cones are over-tilted, $|g|>v_F$, the direction of propagation of the zeroth Landau level of one of the two nodes, but not both, is reversed. The one-dimensional chirality of one zeroth Landau level has thus flipped, despite the fact that the chiral charge of the original Weyl node has not changed. These two statements can be reconciled by realizing that the one-dimensional chirality is defined as the direction of propagation of the zeroth Landau level at momenta close to the Weyl node. As shown in Fig. \ref{fig:WeylEkxB}, a chirality flip requires the zeroth Landau level to develop an S-like shape within a larger momentum range, crossing the Fermi level three times instead of one. One can then define a one-dimensional chirality for the states close to each crossing point. The correspondence with the original chirality $C$ of the original Weyl node is determined by the net chirality of these three sets of states. Because the two additional crossings of the zeroth Landau level with the Fermi energy occur in a momentum range where the zeroth Landau level is merged with bulk Landau levels, the chirality flip as defined above has physical consequences that we will discuss below in section \ref{sec:ProbingOC} and as was predicted before in Ref. \cite{Udagawa2016,Yu2016}.
\begin{figure}[t!]
\captionsetup[figure]{}
\captionsetup[subfigure]{labelformat=empty}
\captionsetup[subfloat]{farskip=0pt,captionskip=1pt}
\centering
\subfloat[]{\includegraphics[width=0.5\columnwidth]{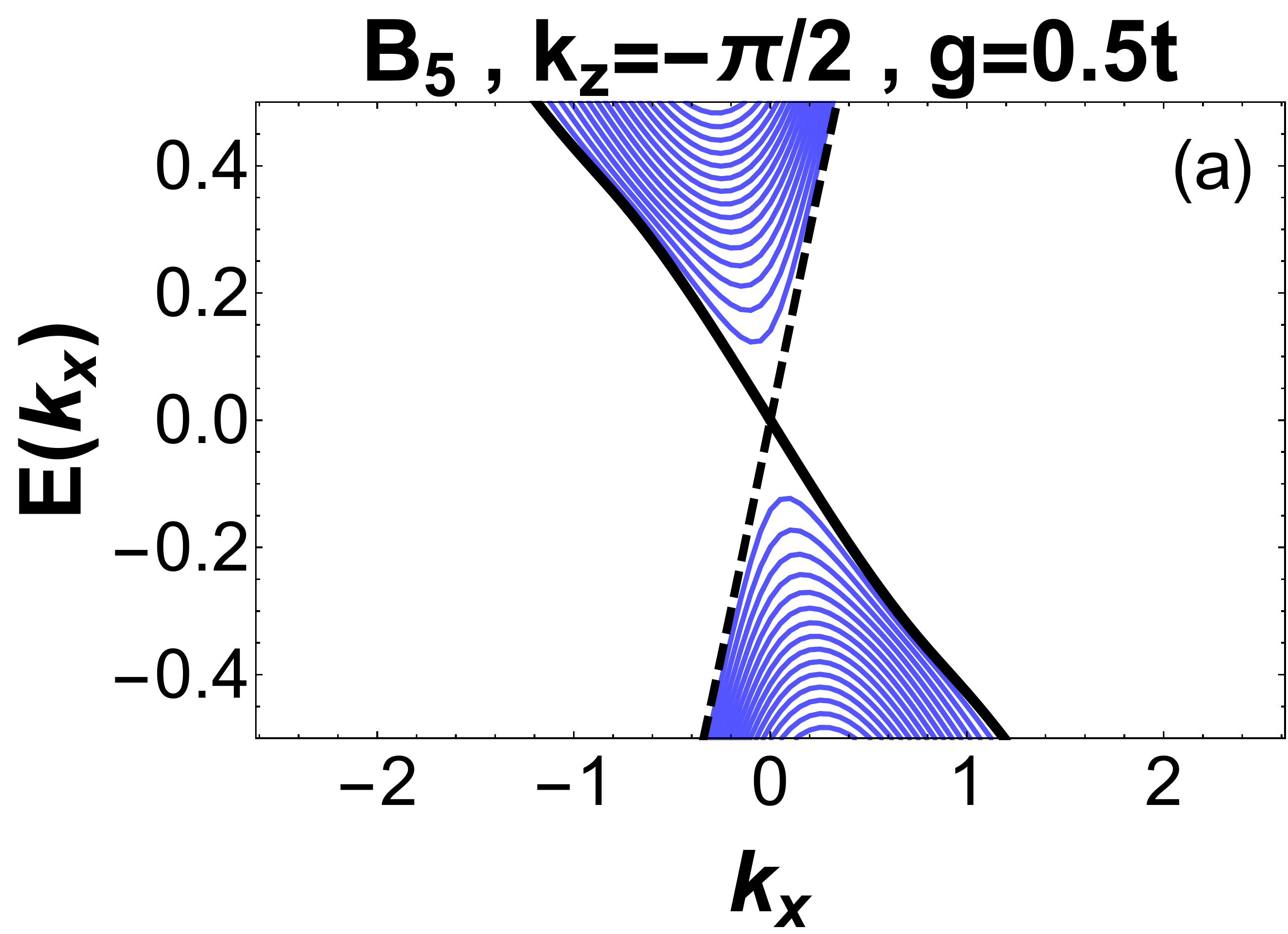}}\hfill
\subfloat[]{\includegraphics[width=0.5\columnwidth]{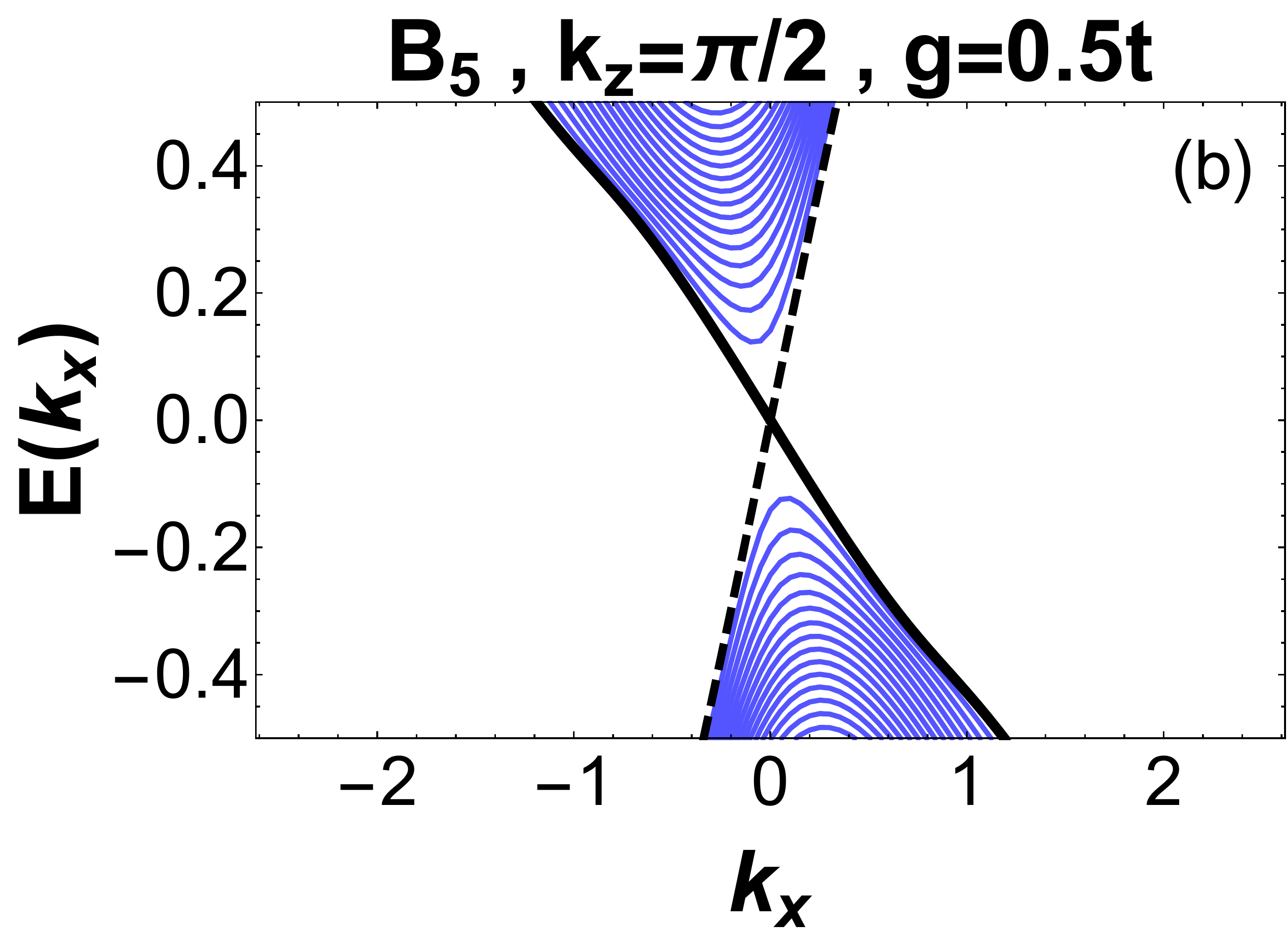}}\hfill
\subfloat[]{\includegraphics[width=0.5\columnwidth]{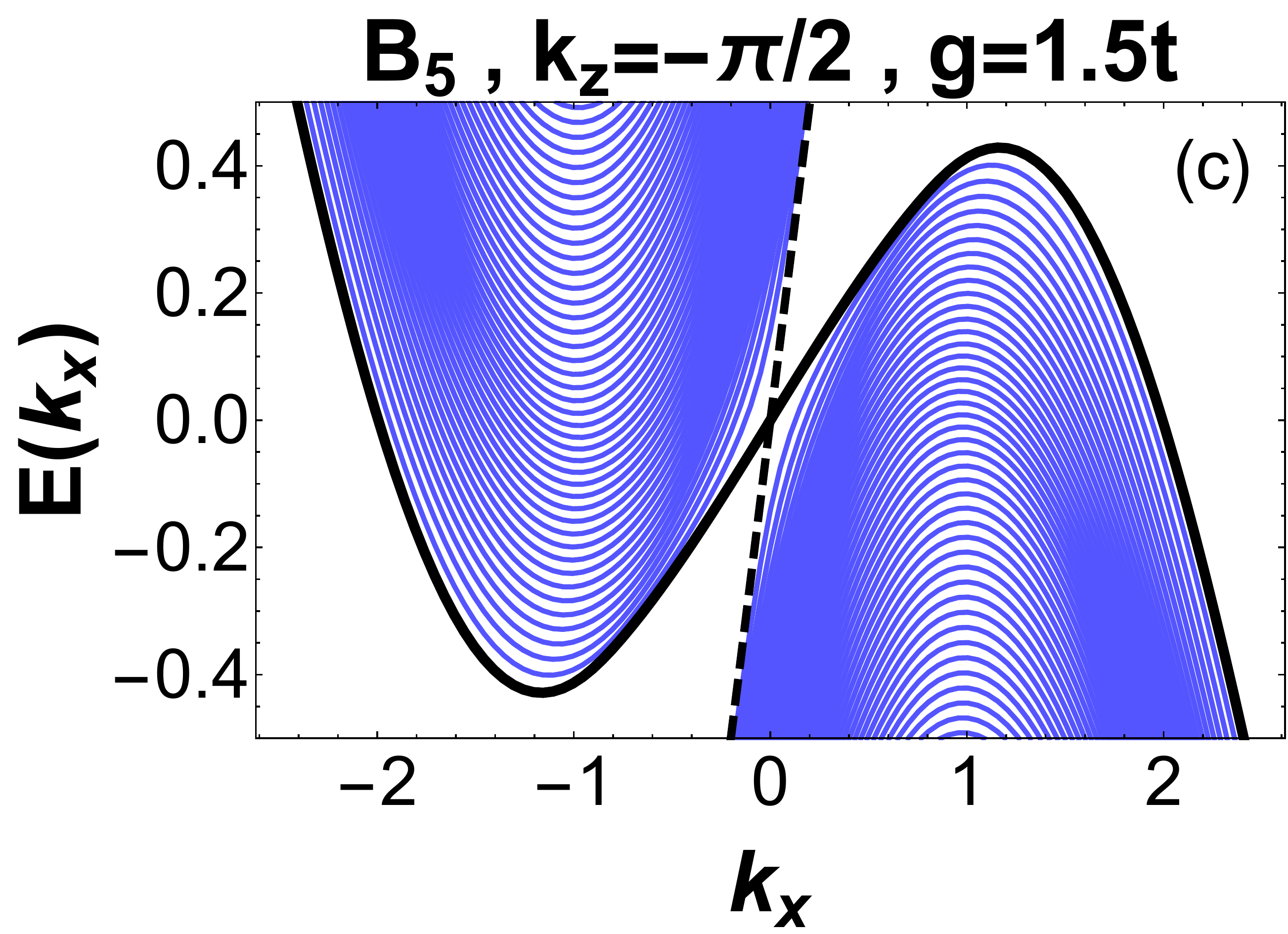}}\hfill
\subfloat[]{\includegraphics[width=0.5\columnwidth]{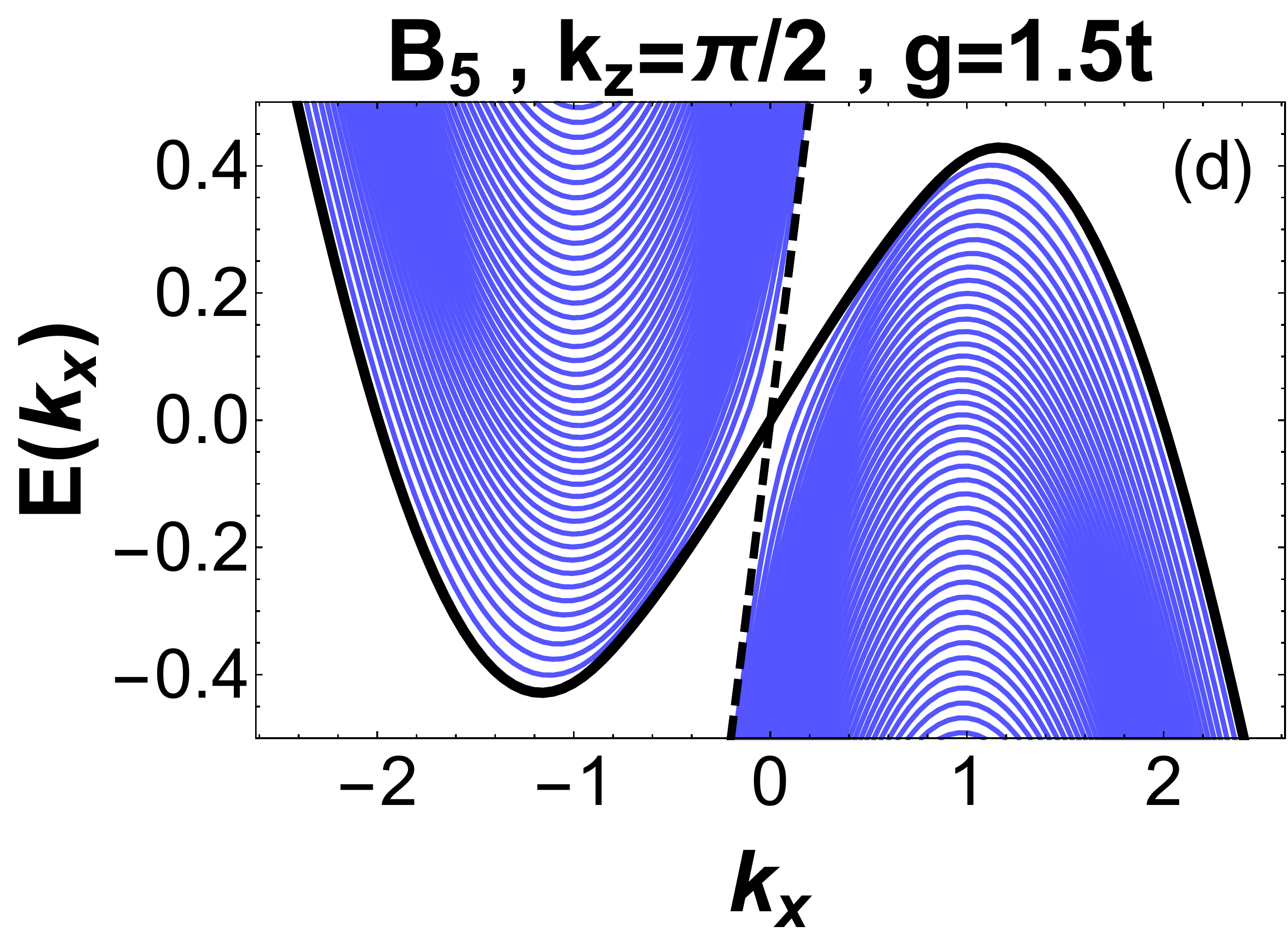}}\hfill
\vspace{-0.2in}\caption{$E(k_{x})$ Spectrum of a Weyl semimetal given by Eq.~$\ref{eq:TBModel}$ in the presence of a constant pseudo-magnetic field $B_5=0.01\hat{x}$. Here, $a=1$, unperturbed node location is at $k_{0}=\pi/2-B_5 n/2$, $n=101$ and different $g$ values. $n\ne0$ bulk modes are colored $\boldsymbol{blue}$, $n=0$ bulk modes are in $\boldsymbol{solid\:black}$ and surface modes in $\boldsymbol{dashed\:black}$. For (a),(b), $g=0.5t$ and the spectrum evaluated close to $k_{z}=-\pi/2,\pi/2$ respectively. for (c),(d), $g=1.5t$ and the spectrum is evaluated at $k_{z}=-\pi/2,\pi/2$ respectively. For the same magnitude of tilt as in Fig.~\ref{fig:WeylEkxB} (c) and (d), the LLL at both nodes experiences chirality flips.}\label{fig:WeylEkxB5}
\end{figure}
In contrast to external magnetic fields, pseudo-magnetic fields can cause the chiralities of the zeroth Landau levels of both nodes to flip simultaneously. Generated intrinsically, pseudo-fields couple to the low-energy Hamiltonian around the two nodes with an opposite sign. This is best illustrated by considering the total low-energy Hamiltonian of two Weyl nodes of topological charge $C=\pm1$ residing at momenta $C\boldsymbol{k_{0}}(\mathbf{r})$, which reads
\begin{eqnarray}
H_{B_{5}}=C v_F \boldsymbol{\sigma}\cdot\left(-i\boldsymbol{\nabla}-C\boldsymbol{k_{0}}(\textbf{r})\right) + g(-i\partial_{\parallel})\sigma_0 ,
\label{eq:WeylHamiltonianB5}
\end{eqnarray}
where we assume the particular choice of $C=1$ at positive $k_0$ and $C=-1$ at negative $k_0$, as this is consistent with the tight binding model we later present in Eq.  \ref{eq:TBModel}. In general, the topological charges can be opposite to this choice, which will lead to an extra minus sign in front of  $C\boldsymbol{k_{0}}$ in Eq. \ref{eq:WeylHamiltonianB5}. The Weyl node separation enters the Hamiltonian similar to the vector potential in Eq.~\eqref{eq:WeylHamiltonianB}, albeit with opposite signs for the two nodes. The curl of the pseudo-vector potential can consequently be understood as the equivalent of a magnetic field $\boldsymbol{B_{5}}=\frac{1}{|e|}\boldsymbol{\nabla}\times\boldsymbol{k_{0}}$ that couples with opposite sign to the two nodes. Close to a given node, such a field is mathematically indistinguishable from a real magnetic field, and the system must consequently develop a spectrum similar to the Landau levels discussed in Eq.~\ref{eq:WeylSpectrumnB}. Indeed, the $n>0$ Landau levels are identical to Eq.~\ref{eq:WeylSpectrumnB}. 
The chiral $n=0$ pseudo-Landau levels, on the other hand, becomes
\begin{eqnarray}
E_{0,C}\left(k_{\parallel}\right)=(\text{sgn}\left(B_{5}\right)v_F+g)k_{\parallel} .
\label{eq:WeylSpectrum0B5}
\end{eqnarray} We observe that if the tilt is strong enough, the chiralities of the zeroth Landau levels can still be flipped, but the flip now occurs simultaneously in the zeroth Landau levels of both nodes, as shown in Fig.~\ref{fig:WeylEkxB5}.

The discussion of the spectral properties close to individual Weyl nodes leads us to conclude that the effect of magnetic fields and pseudo-fields in un-tilted Weyl nodes is difficult to distinguish by any probe without a spectral resolution. A measurement of the optical conductivity will for example be sensitive to the formation of Landau levels, but will not be able to determine if they are caused by real fields or pseudo-fields. Strong tilts provide an effective means to distinguish real magnetic fields from pseudo-fields. If the nodes are over-tilted, $|g|>v_F$, a reversal of a real magnetic field leaves most probes untouched, since a reversal of the magnetic field merely changes the roles of the zeroth Landau levels at the two nodes: the thus far un-flipped Landau level is flipped, while the flipped Landau level is un-flipped. As will be discussed in section \ref{sec:ProbingOC}, these two situations cannot be distinguished by the optical conductivity. In stark contrast, a reversal of the pseudo-field changes the bulk spectrum from having two flipped zeroth Landau levels to having two un-flipped ones, which has a clear signature in the optical conductivity.

\subsection{Tight binding model}
\label{subsec:TB}

Before analyzing potential probes of lowest Landau level chirality flips, we first show that the conclusions of the previous section dealing with a low-energy continuum model also hold from the perspective of lattice models. Although the rationale can be clearly stated based on the low energy theory, it is crucial to take into consideration the full bandstructure for the purpose of keeping track of all contributions coming from other regions of the bandstructure, and in particular, compensation mechanisms enforced by bulk-boundary correspondence. This is especially important when dealing with intrinsic fields, since these are allowed to exist only within the material and hence their existence must leave hallmarks at the surface \cite{Grushin2016,Pikulin2016,Ilan2019}. 

\begin{figure}[t]
\captionsetup[subfigure]{labelformat=empty}
\captionsetup[subfloat]{farskip=0pt,captionskip=1pt}
\subfloat[]{\includegraphics[width=0.8\columnwidth]{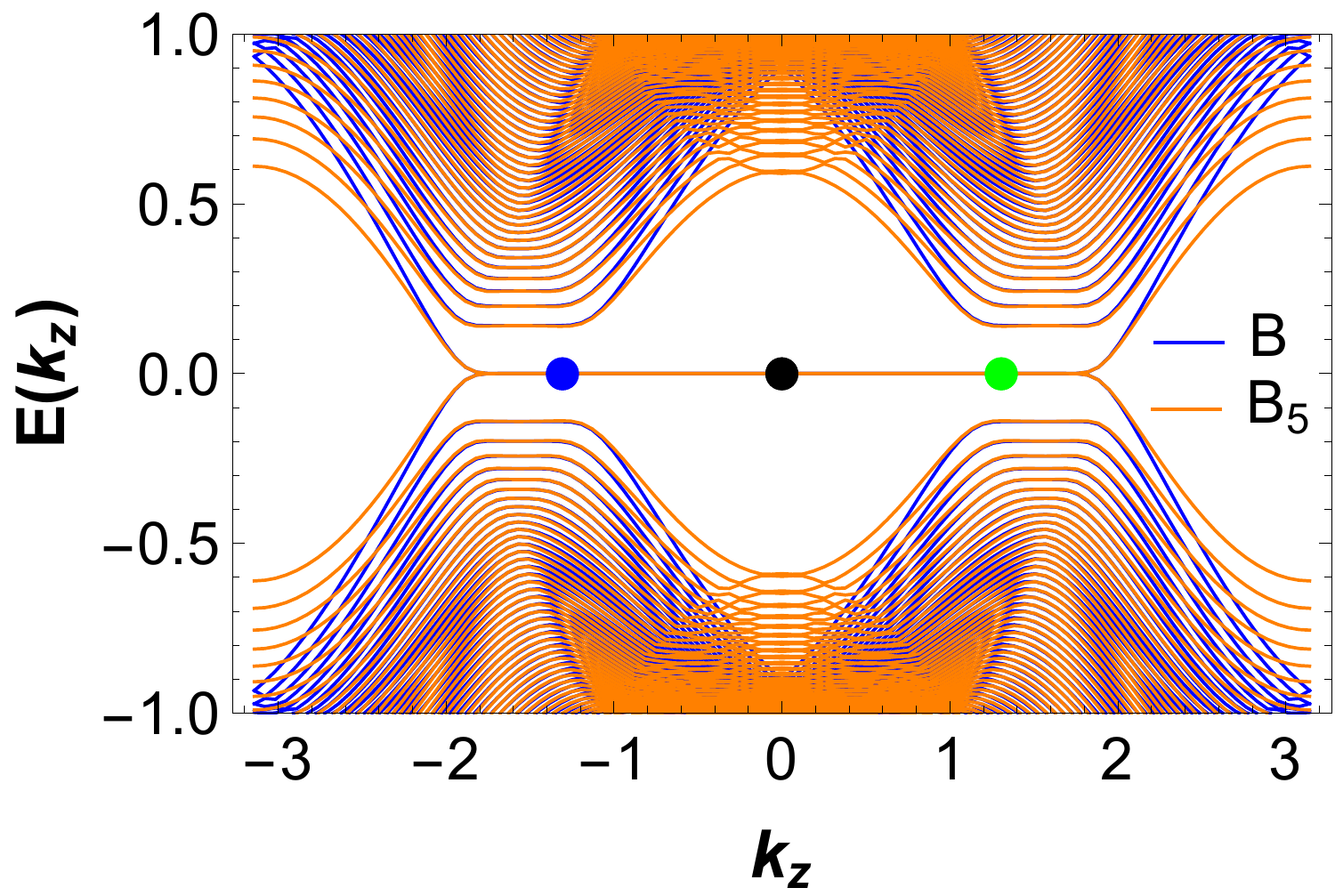}}\hfill
\vspace{-0.2in}\caption{Spectrum of a Weyl semimetals given by Eq.~$\ref{eq:TBModel}$ in the presence of a constant magnetic field $B=0.01\hat{x}$ with with  $M=2$ ($\boldsymbol{blue}$) and a constant pseudo-magnetic field $B_5=0.01\hat{x}$ with unperturbed node position $k_{0}=\pi/2-B_5 n/2$ ($\boldsymbol{orange}$). For both cases $a=1$ and $n=101$. In the vicinity of the Weyl nodes, the fields create flat coinciding Landau levels. The colored dots correspond to the momenta mentioned in Fig. \ref{fig:WeylPseudoMagneticWF} where we study the wavefunctions.}\label{fig:WeylEkz}
\end{figure}

Our first step towards simulating a realistic system is to consider a lattice model of finite size. In finite systems, Weyl semimetals support topological surface states whose Fermi lines are known to form open Fermi arcs ending at the projections of the Weyl nodes onto the respective surface. Because pseudo-fields also affect these surface states, a model for pseudo-field effects can only be complete if it includes the surfaces as well. One can for example show that a linear spatial gradient of the Weyl node separation in the bulk leads to Fermi arcs of different lengths at two opposite surfaces \cite{Grushin2016}. In addition, the fermion doubling theorem \cite{Nielsen1981} guarantees that a lattice model can never have a net chirality, including the (pseudo-) Landau levels of three-dimensional systems. Correspondingly, 
the co-propagating chiral zeroth Landau levels in the bulk of a Weyl semimetal subject to a pseudo-magnetic field are compensated by two modes on the surfaces. This form of bulk-edge correspondence suggests that any claims made with respect to chiral structures in the bulk, and in particular their probes, should be carefully weighted along with potentially compensating effects coming from the surface states. 

To make the comparison with previous works more tractable, we focus on the two band model used in Ref. \cite{Udagawa2016}
\begin{eqnarray}
H_{2b}(k)&=&t[\sin(k_xa)\sigma_x+\sin(k_ya)\sigma_y+\cos(k_za)\sigma_z]+\label{eq:TBModel}\\ 
&&m[M-\cos(k_xa)-\cos(k_ya)]\sigma_z- g\sin(kx)\sigma_0 ,\nonumber
\end{eqnarray}
where we work in units of $e = \hbar = 1$ throughout. This model exhibits one pair of Weyl nodes at $\boldsymbol{k_{0}}=\left(0,0,\pm\cos^{-1}\left(M-2\right)\right)$  when $1<\left|M\right|<3$, two pairs of Weyl nodes for $\left|M\right|<1$, and corresponds to a fully gapped insulator for $\left|M\right|>3$. In the remainder, we focus on the regime $1<\left|M\right|<3$, and in this subsection choose $t=-m=-1$. We consider a slab with $n$ unit cells in $\hat{y}$-direction, and  periodic boundary conditions along $\hat{x}$ and $\hat{z}$. Adding a magnetic field by minimal coupling to $\boldsymbol{A}=\left(0,0,-By\right)$, Figs. \ref{fig:WeylEkz} and \ref{fig:WeylEkxB} depict the resulting Landau level spectrum. At momenta close to the nodes, the spectrum of the full slab reproduces the Landau level of the low-energy model discussed above. Away from the nodes, the levels smoothly interpolate between the two nodes.

\begin{figure}[t]
\captionsetup[figure]{}
\captionsetup[subfigure]{labelformat=empty}
\captionsetup[subfloat]{farskip=0pt,captionskip=-2pt}
\centering
\subfloat[]{\includegraphics[width=0.5\columnwidth]{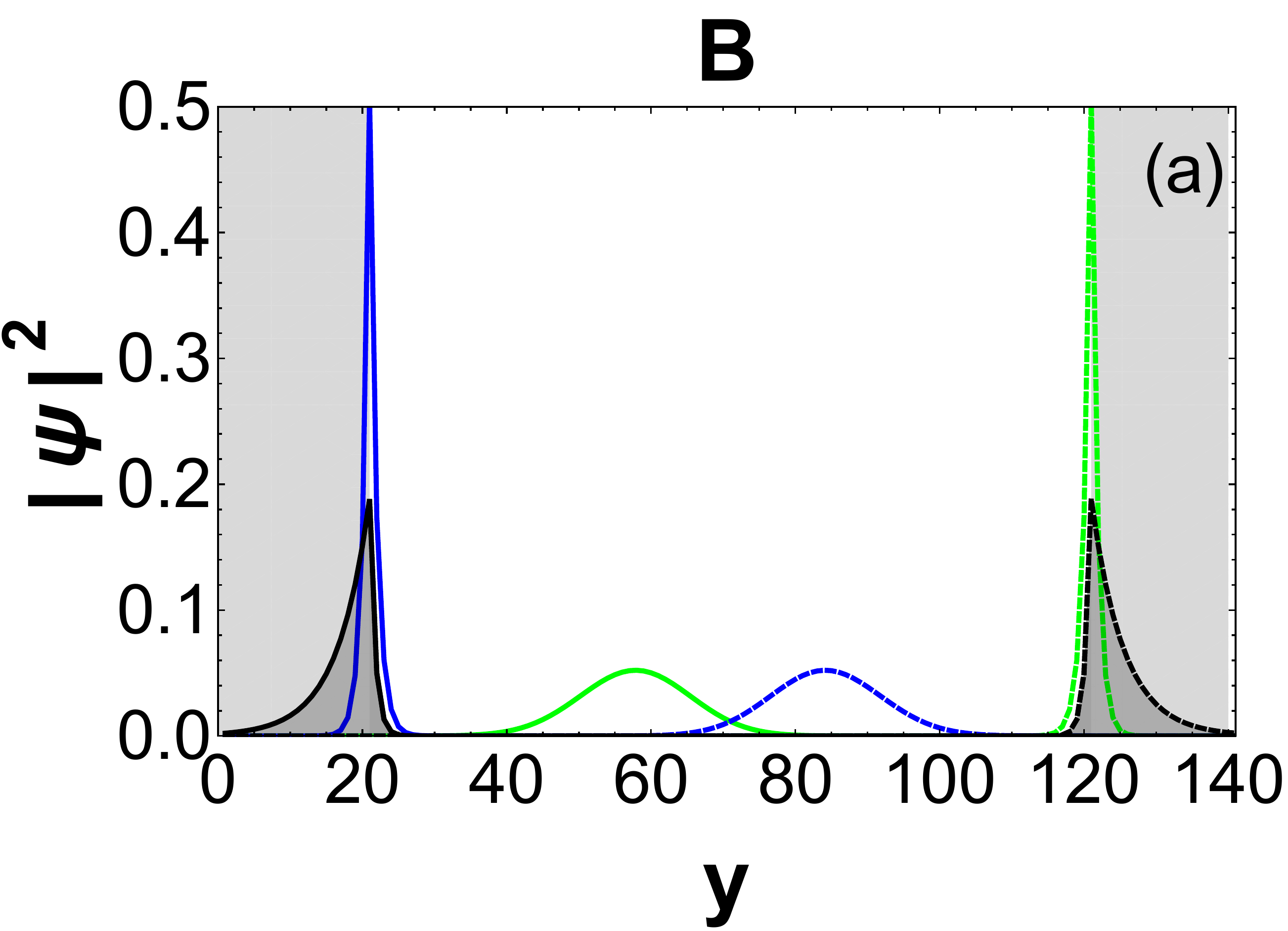}}\hfill
\subfloat[]{\includegraphics[width=0.5\columnwidth]{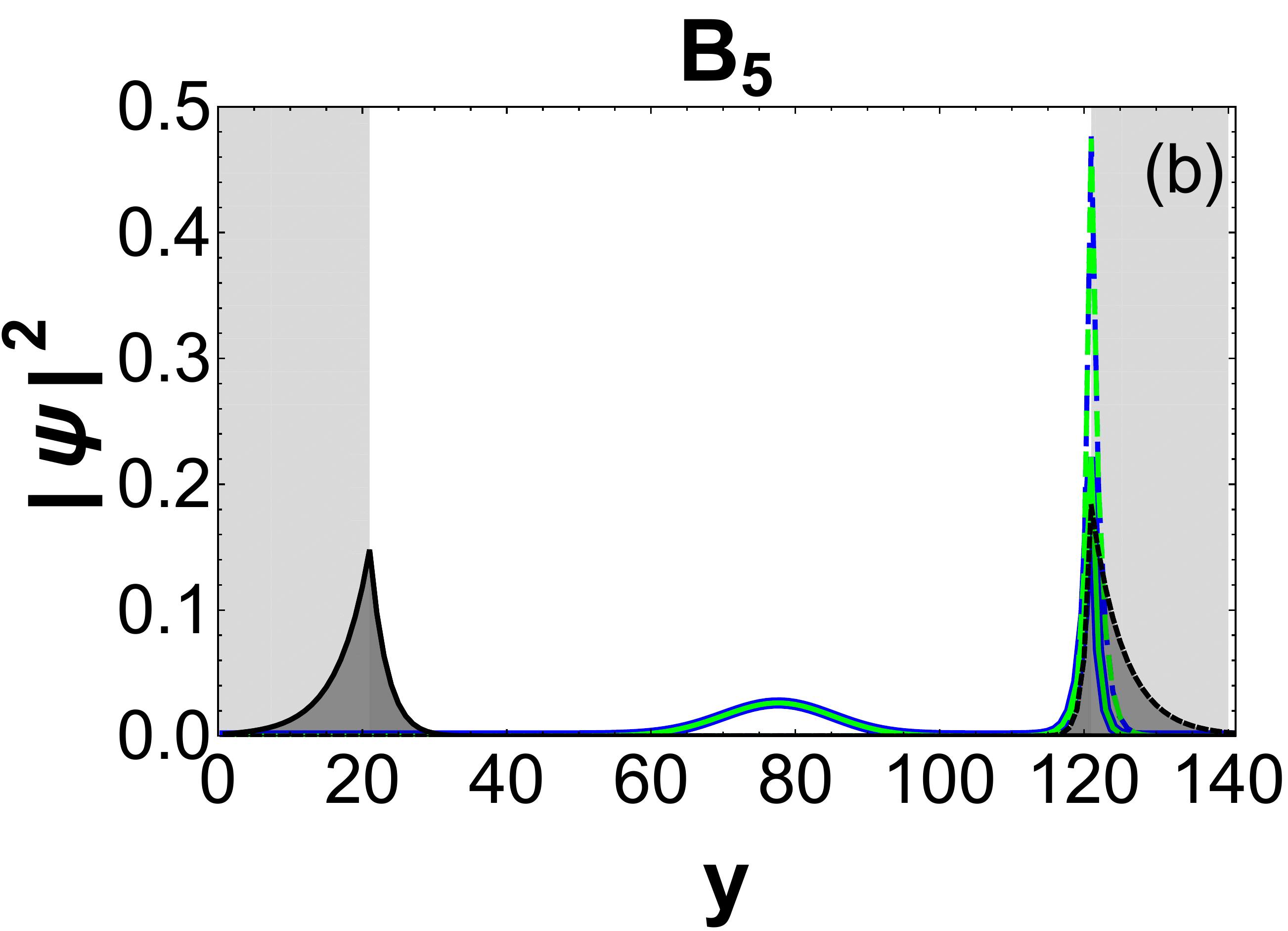}}\hfill
\vspace{-0.1in}\caption{\label{fig:WeylPseudoMagneticWF} Wave functions of the two zero energy bands of a Weyl semimetals in the presence of a $B$ or a $\mathbf{B}_5$. The states belong to the $k_{x}=0$ band plotted for different $k_z$, as indicated in Fig. \ref{fig:WeylEkz}. A full or dashed line differentiates between the doubly degenerate zero bands. Green and blue curves in (b) completely coincide, exemplifying that the orbitals behave the same at the two nodes in the presence of a $\mathbf{B}_5$. Gray regions represent trivial insulators introduced by setting $M=3$ in Eq. \ref{eq:TBModel} and adding $H_{gap}=Y\sigma^{y}$ everywhere. In the metallic region $M=2$, $Y=0.5$, $n_{Weyl}=101$, $n_{insulator}=20$,  $k_{0}=\pi/2$. Field values are  $B=0.01$, $B_5=0.01$.}
\end{figure}

We confirm the chiral structure of Landau levels and surface states by studying the wavefunctions of the bands at zero energy. The surface states forming a dispersion-less band in $k_z$ between the Weyl nodes remain largely unaffected as long as there are no bulk Landau levels nearby in energy. This is best highlighted by studying the wave functions of the two bands for $k_x=0$ at different momenta $k_z$. Fig. \ref{fig:WeylPseudoMagneticWF} shows that the states in the (nearly) two-fold degenerate zero energy bands remain localized at one of the two boundaries of the slab around $k_z=0$, where no bulk states is close-by in energy. Closer to one of the nodes, we find that one of these two states is localized at the surface, while the other is spread out into the bulk. A similar behavior is observed close to the other node, but with switched roles  of the localized and spread-out bulk state. 

To contrast the effect of a real magnetic field with pseudo-magnetic fields, we now allow $M$ to vary in position such that a uniform pseudo-magnetic field $\boldsymbol{B_{5}}=\nabla\times\left(B_5 y\hat{z}\right)=B_5\hat{x}$ in created in the bulk. The corresponding spectrum is plotted in Figs. \ref{fig:WeylEkxB5} and \ref{fig:WeylEkz}. As expected, we find that the $n\ge1$ band structure is similar to the Landau levels of an external magnetic field close to the nodes. However, the spectrum of the $n=0$ chiral modes is different, as well as the orbital character of these states. This is illustrated in Fig. \ref{fig:WeylPseudoMagneticWF} depicting the wave functions of the zero energy states $k_{x}=0$ for different values of $k_{z}$. In contrast to what happens for $B$, for $B_5$ the states spreading into the bulk in the vicinity of both nodes belong to the same band, while states from the other band are both confined to the same surface. This reflects that the bulk chiral modes at both nodes have an identical dispersion, as predicted by Eq. \ref{eq:WeylSpectrum0B5}.

Next, an additional tilt of the nodes parallel to the external or pseudo-magnetic fields (the $\hat{x}$-direction) is generated by the term $g\sin\left(k_{x}a\right)\sigma_{0}$. As shown in Figs. \ref{fig:WeylEkxB} and \ref{fig:WeylEkxB5}, the tight-binding model now reproduces perfectly the effects of a tilt discussed in the low-energy model. For an external magnetic field $\mathbf{B}$, a tilt changes the slopes of the counter-propagating bulk Landau level such that values $|g|>|t|$ correspond to a band structure in which one of the two bulk zeroth Landau levels has a flipped chirality. In contrast, the Landau levels generated by a pseudo-magnetic field $\mathbf{B}_5$ are co-propagating. In the over-tilted regime $|g|>|t|$, both bulk zeroth Landau levels are hence simultaneously flipped, or not. 

It is important to stress that when there is no chirality flip in the bulk, an S-shape mode may still appear, but will correspond to the surface state experiencing the chirality flip. This will happen in cases where the tilt direction is perpendicular to the direction of the Weyl node separation, as considered in the above example. This should be expected since it has been pointed out in Ref.~\cite{Grushin2016}, that Fermi arcs can be understood as a lowest Landau levels of a strong $\mathbf{B}_5$ confined to the surface plane, resulting from the annihilation of Weyl points due to the lattice termination.

\section{Probing over-tiled cones: optical conductivity in the presence of pseudo-fields}
\label{sec:ProbingOC}

While any probe that maps the band structure of a Weyl semimetal might in principle identify the presence of a type II phase, such a measurement is in practice harder than it seems. Probing the bulk and surface states require laser sources in the UV to deep UV frequencies, respectively. Even if signatures for over-tilted cones would be imprinted in the surface states, it is generally challenging to resolve and detect the Fermi arcs due to the existence of finite particle and hole pockets in their vicinity, or their shortness in common materials~\cite{Ma2019}. Thermodynamic probes, such as the specific heat, will of course be able to resolve the transition between a Weyl semimetal of type I with tiny Fermi pockets around the nodes, to a type II semimetal with extended Fermi pockets. Distinguishing a material with trivial electron and hole pockets from a Weyl semimetal of type II, however, is not easily done with thermodynamics. A similar problem arises when discussing the transport phenomenon of an anisotropic negative longitudinal magneto-resistance, which is considered to be a strong evidence for a type II Weyl phase, contributed to the chiral anomaly. It was found that other systems may also exhibit such behaviour, and on the other hand, in some cases type II Weyl semimetals may exhibit isotropic negative longitudinal magneto-resistance  \cite{Hu2019,Son2013,Knoll2019,Reis2016,Liang2018,Armitage2018}. As we discuss now, it is not necessary to experimentally map the full band structure to detect a type II phase in a Weyl semimetal: despite being a probe that averages in momentum space, the optical conductivity is enough to fingerprint a type II phase in a Weyl semimetal, and to identify the direction of the tilt.

The optical conductivity experimentally corresponds to exciting electron-hole pairs by the absorption of a photon of frequency $\omega$. Because the velocity of light is much larger than the Fermi velocity, these particle-hole pairs carry a vanishing momentum on the scale of the Brillouin zone, and electrons are essentially excited vertically in the band structure. We evaluate the optical conductivity using the Kubo formula \cite{Allen2006} as
\begin{eqnarray}
\sigma_{xx}\left(\omega\right)=\int \frac{d^2k}{iV}\sum_{n \ne m}\frac{f(\varepsilon_{n})-f(\varepsilon_{m})}{\varepsilon_{n}-\varepsilon_{m}} \frac{|\left\langle n\left|v_{x}\right|m\right\rangle|^2 }{\omega+\frac{i}{\tau}+(\varepsilon_{n}-\varepsilon_{m})},\quad
\label{eq:OC_Kubo}
\end{eqnarray}
where $V$ is the volume of the system, $d^2k=dk_xdk_z$,  $\varepsilon_{n}$ are eigenenergies, $|n\rangle$ are the corresponding eigenstates, $f(\varepsilon_{n})=1/(1+\exp(\beta(\varepsilon_{n}-\mu)))$ is the Fermi-Dirac distribution with inverse temperature $\beta$ and a chemical potential $\mu$, and $v_{x}=\partial H/\partial k_{x}$ is the velocity operator. To qualitatively take into account the effect of impurity scattering, we follow Ref.~\cite{Udagawa2016}, and modify the electronic propagator by the ad-hoc introduction of a finite life-time $\tau$. To allow the direct comparison of our results with the ones of  Ref. \cite{Udagawa2016}, we use the same set of parameters,  $\mu=0$, $\beta=1/0.01t$ and $\tau^{-1}=0.1t$. We then compute the optical conductivity for the tight binding model in Eq. \ref{eq:TBModel} with a system of length $n=101$, open boundary conditions, and  $t=0.2$, $m=3t$ and $a=1$.

\begin{figure}[t]
\captionsetup[subfigure]{labelformat=empty}
\captionsetup[subfloat]{farskip=0pt,captionskip=1pt}
\subfloat[]{\includegraphics[width=0.5\columnwidth]{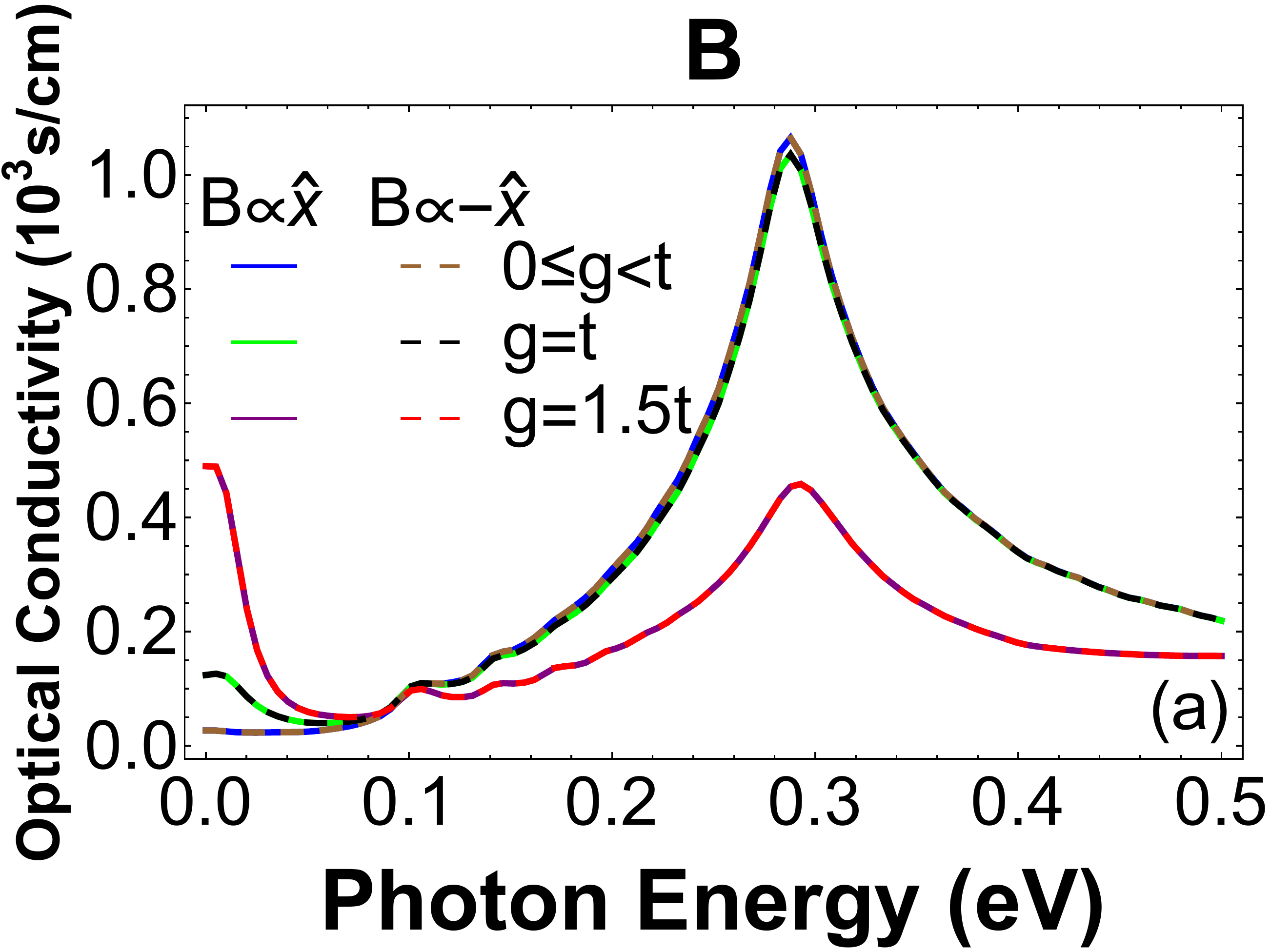}}\hfill
\subfloat[]{\includegraphics[width=0.495\columnwidth]{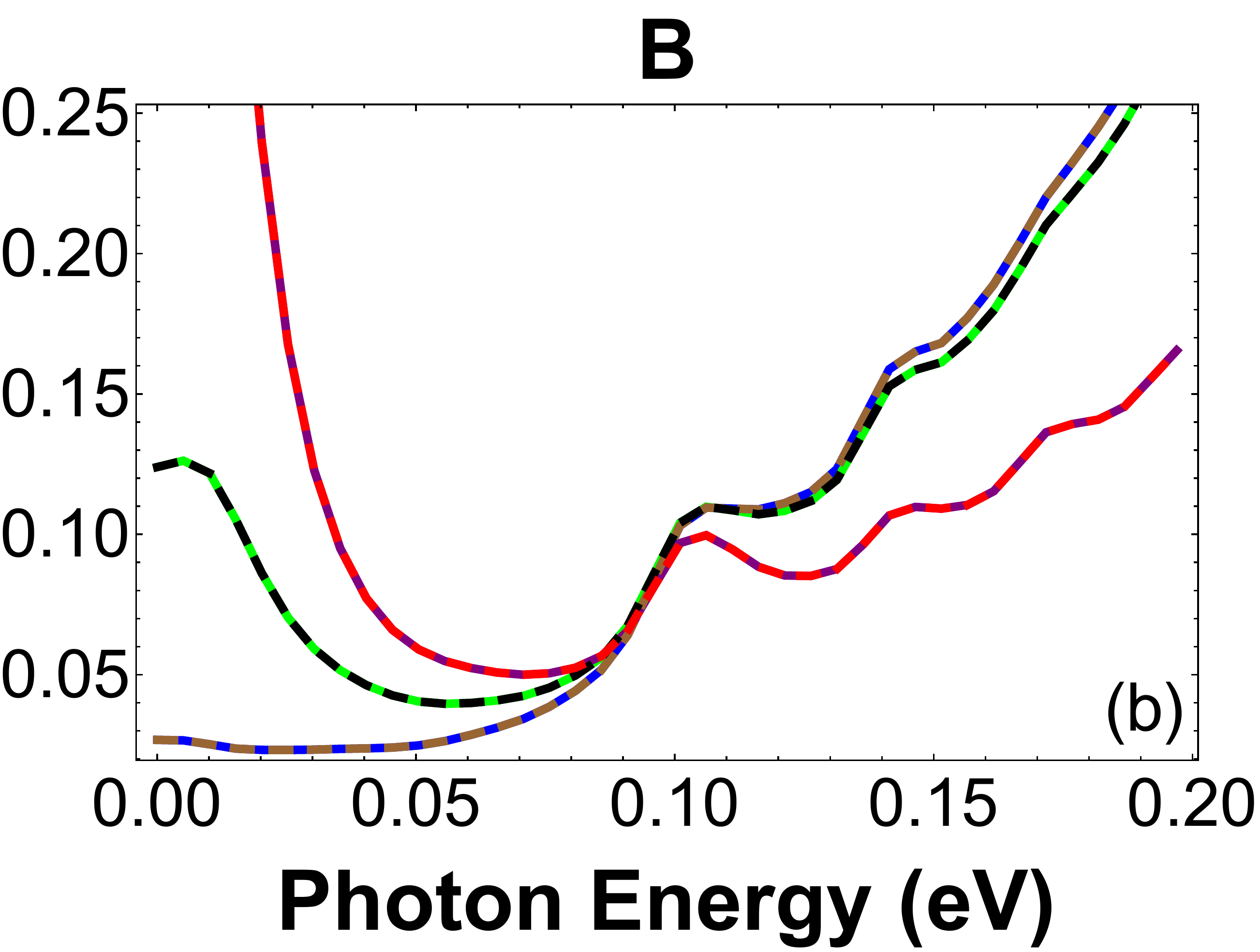}}

\subfloat[]{\includegraphics[width=0.5\columnwidth]{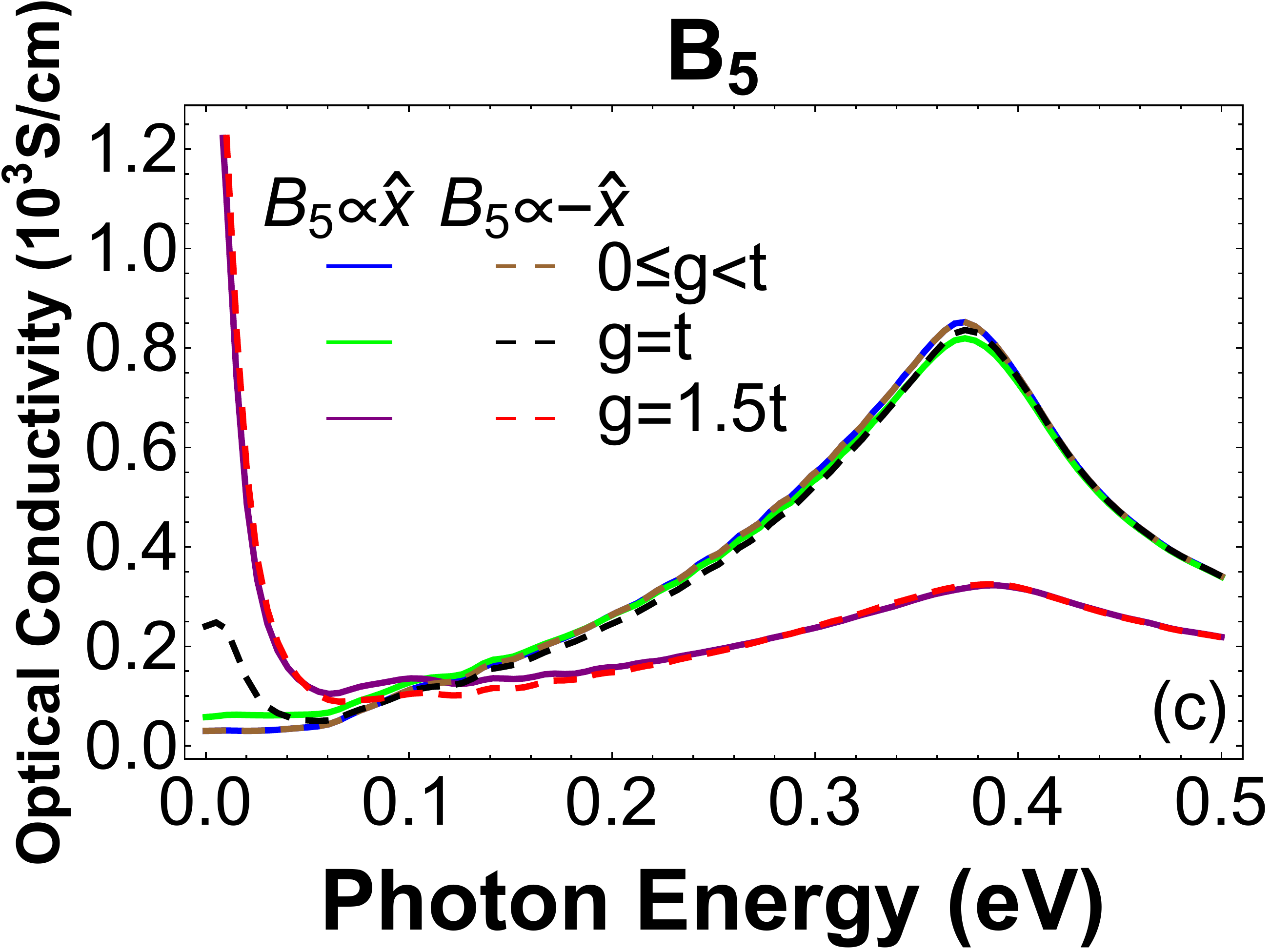}}\hfill
\subfloat[]{\includegraphics[width=0.495\columnwidth]{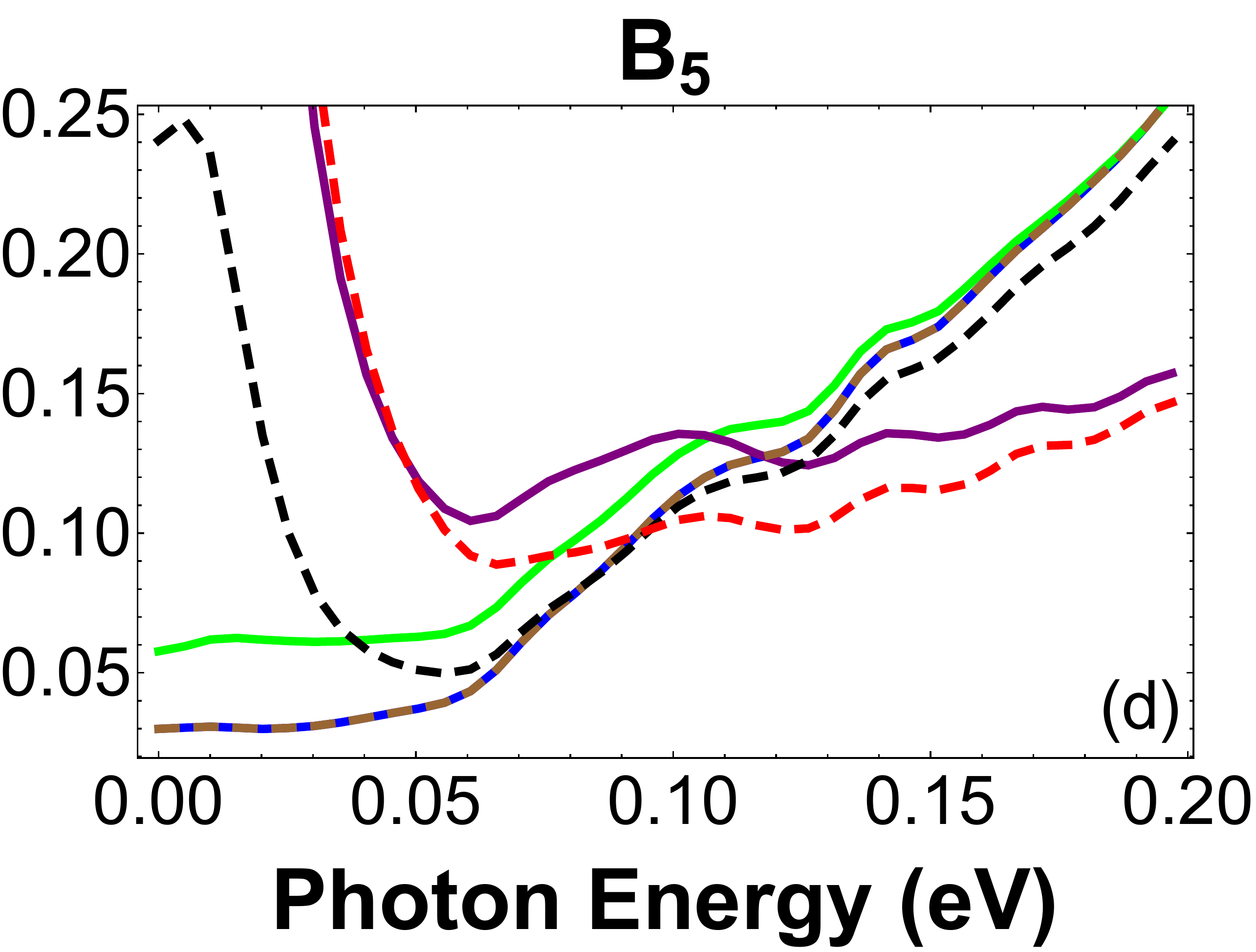}}
\vspace{-0.1in}\caption{ Optical conductivity for a type II semimetal described by eq. \ref{eq:TBModel} with the parameters in the main text, and with $|B|,|B_5|=\pi/(n-1)$ in (a) and (c) respectively, for different values of tilt $g$ and system size $n=101$. For solid curves the field is in the $\hat{x}$ direction, while dashed curves the field is in $-\hat{x}$ direction. In (b),(d) we zoom in on  the first spectral transitions of (a) and (c) appearing as oscillations.}\label{fig:OC_BandB5} 
\end{figure}

For periodic boundary conditions, the optical conductivity of a Weyl semimetal has peaks at characteristic frequencies that fingerprint the Landau levels in an external magnetic field for all values of the tilt~\cite{Udagawa2016}. With open boundary conditions the oscillations are qualitatively similar. Importantly, we find that the optical conductivity is independent of the direction of the external magnetic field, as evidenced by the upper plots of Fig.~\ref{fig:OC_BandB5}. Namely, in agreement with our above discussion, a reversal of the direction of the field reverses the velocity of the two zeroth Landau levels simultaneously. Because of its momentum averaging nature, the optical conductivity cannot distinguish the two valleys, and hence cannot detect this change. 

While the optical conductivity exhibits oscillations with peaks at the same positions for a pseudo-magnetic field, see  Fig.~\ref{fig:OC_BandB5}, the height of these peaks is sensitive to the direction of the field. This is demonstrated in the lower plots of  Fig.~\ref{fig:OC_BandB5}, and is in stark contrast to the effect of a reversal of the direction of an external magnetic field, see upper plots of  Fig.~\ref{fig:OC_BandB5}. More precisely, while the optical conductivity is invariant to a change in the pseudo-field direction in the type I-regime, it is sensitive to the direction of the field in the type II-regime. The differences in optical conductivity between the two field directions is found to be strongest at intermediate frequencies just above the large peak at $\omega=0$, which is contributed by the spectral transitions between the chiral landau level and the first few $n>0$ levels. 

The sensitivity of the optical conductivity to a reversal of a pseudo-field in the over-tilted regime is a direct consequence of the chirality flips occurring for $|g|>|t|$, and is a central result of this work. As discussed above, a sign-reversal of a real magnetic field in the over-tilted regime merely changes which of the two bulk Weyl nodes exhibits a chirality-flipped zeroth Landau level. A pseudo-field reversal, on the contrary, changes the bulk band-structure from having two chirality-flipped zeroth Landau levels to having none, see Eqs.~\ref{eq:WeylSpectrum0B} and \ref{eq:WeylSpectrum0B5}. Optical transitions of a single Weyl node in an external magnetic field transitioning between type I and II were discussed in Ref.~\cite{Yu2016}. 

Our discussion specializes on fields nearly parallel or anti-parallel to the tilt direction.  Ref.~\cite{Soluyanov2015} has shown that for the LLL to persist in type 2 semimetals, the angle between  the direction of the tilt and the field needs to be lower than a critical angle defined by the magnitude of the tilt. Since the zeroth Landau level is crucial for the our result to hold, we expect the effects to vanish beyond such critical angles, consistent with the findings of ref.~\cite{Udagawa2016} that demonstrated a smoothening of the oscillations in the optical conductivity in the presence of a magnetic field perpendicular to the tilt direction.

To summarize, our detailed analysis shows that the optical conductivity is a sensitive detector for the type II regime in Weyl semimetals. In particular, we find that when Weyl node host bulk chiral Landau levels with the same dispersion induced by a pseudo-field, co-tilted nodes in the type II regime result in either a simultaneous chirality flip, or no chirality flip at all, depending on the direction of the pseudo-field with respect to the tilt. The optical conductivity is sensitive to the direction of the pseudo-field with respect to the tilt direction: a simultaneous chirality flip leads to a lower optical conductivity at the energies of the transitions between the zeroth Landau level and the $n>0$ levels. This asymmetry in the field direction would reveal the direction of the tilt. We also find that this measure is robust, to some extent, to inaccuracies in the field strength when reversing the direction. This was tested by simulating systems with different pseudo-field strengths and observing that the hierarchy between the curves of oppositely directed fields remains for a range of field strengths.  

\section{Counter tilted cones vs. co-tilted cones: switching the role of $B$ and $B_5$}
In the preceding sections we considered a co-tilted two-node system. Nevertheless, our result can be straightforwardly generalized to counter-tilted cones. The intermediate state can then be easily inferred. 

Consider a two node system where the nodes are over-tilted but in opposite direction (i.e. counter-tilted). A pseudo-magnetic field, acting with an opposite sign at the two nodes, would then result, according to Eq.~\ref{eq:WeylSpectrum0B5}, in a chirality flip of the lowest Landau level of one node only. Namely, it now acts similar to a magnetic field in the case of co-tilted cones. This does not lead to an asymmetry in measuring the optical conductivity with respect to the field direction. 

On the contrary, Landau levels created by a magnetic field would lead to either a simultaneous flip, or no flip, depending on the direction of the field. This would lead to an asymmetry of measuring the optical conductivity with respect to the field direction. An optical conductivity measurement in a system with counter-tilted cones therefore exhibits similar phenomena to the ones discussed above, with the roles of magnetic and pseudo-magnetic fields switched (under the assumption that axis of the tilt and the field coincide). We have confirmed this numerically finding results similar to Fig. \ref{fig:OC_BandB5}, which are not shown here. 

To conclude, for co-tilted cones the optical conductivity is invariant to flipping the direction of a magnetic field, but asymmetric with respect to flipping the direction of a pseudo-field, while for counter-tilted cones the converse is true. Naturally, more general scenarios may occur, where the two nodes are not tilted with the same amplitude or precisely in the same direction. Then, we expect that in the type II regime there will be an asymmetry in a measurement of the optical conductivity. Applying, separately, a magnetic or pseudo-magnetic field, and switching their directions, will allow to identify over-tilted cones along with the direction of the tilt.  

\section{Designing type II semimetals with pseudo-fields: layer constructions}
Considering the limited control over the properties of real systems and the large number of nodes known materials exhibit, a manipulation of specific isolated properties such as tilt and Weyl node separation might be challenging in the currently available Weyl semimetals. There are two strategies to tackle this problem. One is to search for more ``ideal'' Weyl semimetals in nature. Indeed, current efforts to systematically identify topological materials give hope that strain-tunable materials realizing our above minimal model with only two Weyl nodes at the Fermi surface, will be available in the near future \cite{Armitage2018,Ma2019,Hu2019}. Alternatively, one can try and push existing materials into a more ideal Weyl semimetal phase by a suitable engineering of their properties. In the following, we discuss two such schemes to engineer Weyl semimetals with a low number of Weyl nodes allowing control of tilt and nodal separation in layered heterostructures. These layered materials can be used to test the prediction of the previous section regrading the optical conductivity as a tool to probe the type of the Weyl semimetal.

\subsection{Strong topological insulator multilayers coupled by spin-dependent hoppings}
\label{sec:STI}

A seminal proposal~\cite{Burkov2011} to create a Weyl semimetal is a 3D superlattice consisting of alternating thin films of 3D magnetic strong topological insulator (TI) and ordinary insulators. In the phase space of the different couplings in the problem, interlayer vs. intralayer, a region in parameter space emerges in which two Weyl points appear in the 3D Brilluion zone. The bulk Hamiltonian is given by
\begin{align}
% \begin{array}{c}
H\left(k\right)=&v_{F}\tau_{z}\left(-\sigma_{y}k_{x}+\sigma_{x}k_{y}\right)+m\sigma_{z}+\Delta_{S}\tau_{x}\nonumber\\
&+\frac{1}{2}\Delta_{D}\tau_{+}e^{ik_{z}d}+\frac{1}{2}\Delta_{D}\tau_{-}e^{-ik_{z}d} ,
% \end{array}
\label{eq:LCHamiltonain}
\end{align}
where $v_F$ is the Fermi velocity of the 2D surface Dirac states of the 3DTI, $\boldsymbol{\sigma}$ and $\boldsymbol{\tau}$ are Pauli matrices acting on the spin and surface degrees of freedom respectively, $\tau_{\pm}=\left(\tau_{x}\pm i\tau_{y}\right)$, $k_x$ and $k_y$ are the momenta in the plane of the layers, $\hat{z}$ is the stacking direction, and $d$ is the double layer width. The parameters $m$, $\Delta_{S}$, and $\Delta_{D}$ represent the magnetic gap within a TI layer, the hopping strength across a TI layer, and the hopping between two TI layers across an insulating layer, respectively.
The resulting spectrum is
\begin{align}
% \begin{array}{c}
E_{\pm}\left(k\right)^{2}=v_{F}^{2}\left(k_{x}^{2}+k_{y}^{2}\right)+\left(m\pm\Delta\left(k_{z}\right)\right)^{2} ,\nonumber\\
\Delta\left(k_{z}\right)\equiv\sqrt{\Delta_{S}^{2}+\Delta_{D}^{2}+2\Delta_{S}\Delta_{D}\cos\left(k_{z}d\right)} .
% \end{array}
\label{eq:LCSpec}
\end{align}
 For the phase space region in which $\left(\Delta_{S}-\Delta_{D}\right)^{2}<m^{2}<\left(\Delta_{S}+\Delta_{D}\right)^{2}$, this model exhibits two topologically stable Weyl nodes separated along the the $k_z$ axis at
\begin{align}
% \begin{array}{c}
k_{0}=\frac{1}{d}\left(\pi\pm\cos^{-1}\left(1-\frac{m^{2}-\left(\Delta_{S}-\Delta_{D}\right)^{2}}{2\Delta_{S}\Delta_{D}}\right)\right) ,
% \end{array}
\label{eq:LCNodeLocation}
\end{align}
This model does not facilitate a tilt of the Weyl nodes.

\begin{figure}[t]
\captionsetup[subfloat]{labelformat=empty}
\captionsetup[subfloat]{farskip=0pt,captionskip=-100pt}
\subfloat[]{\includegraphics[width=0.50\columnwidth]{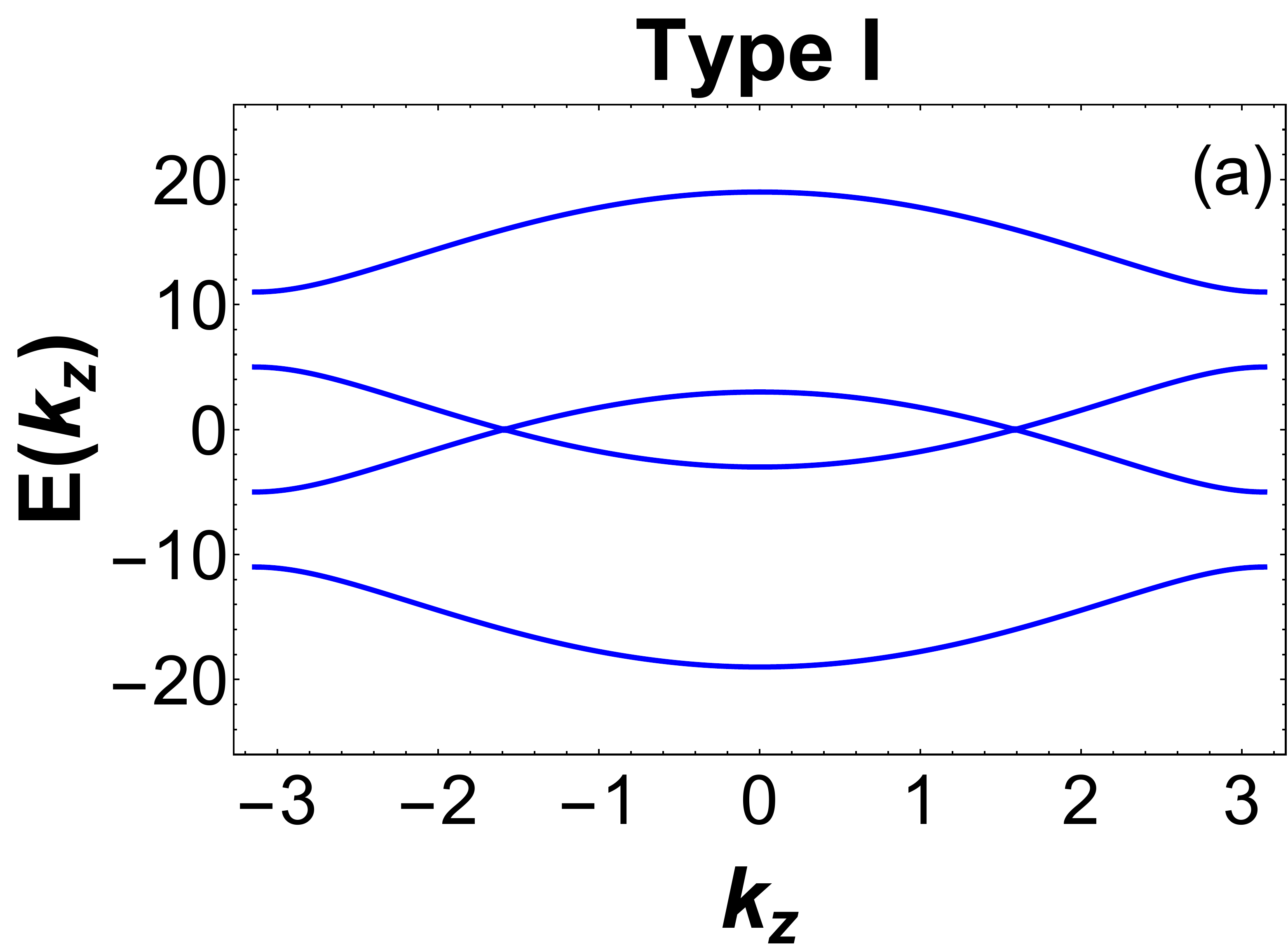}}
\subfloat[]{\includegraphics[width=0.50\columnwidth]{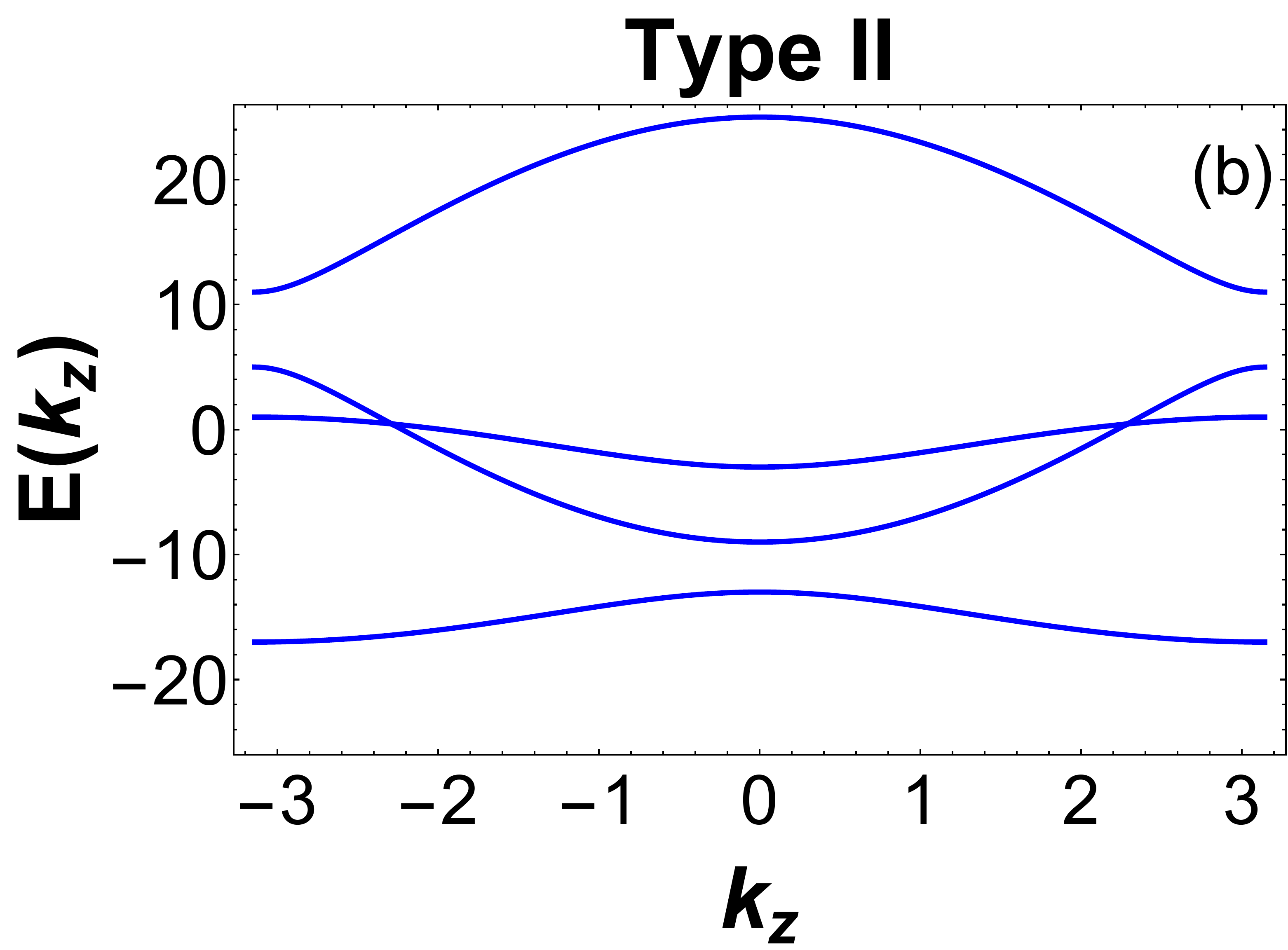}}
\vspace{-0.1in}\caption{\label{fig:LC_Tilted} Band structure of the magnetic TI-trivial insulator superlattice in Eq.~\ref{eq:LCHamiltonain} supplemented by the spin dependent hopping Eq.~\ref{eq:SpinDepTunTerms}. Here $d=1$, $v_{F}=1$, $m=8$, $\Delta_{s}=7$, $\Delta_{d}=4$. The band structures are computed at $k_x=k_y=0$. In the left panel $\Delta_{z}=0$ (type I), and in the right panel $\Delta_{z}=6$ (type II).}
\end{figure}

To allow for a tilt and the creation of a pseudo-field, we change the model by introducing various spin dependent hopping terms across the trivially insulating layers of the form
\begin{eqnarray}
\begin{array}{c}
\Delta_{D}\rightarrow\Delta_{D}+\Delta_{i}\sigma_{i} ,
\end{array}
\label{eq:SpinDepTunTerms}
\end{eqnarray}
where $i\in\left\{x,y,z\right\}$.
A spin-dependent tunneling is, in fact, not a strong requirement: since the topological insulators are required to have time reversal symmetry broken by magnetism, spin-rotation invariance for any process involving these layers is broken to begin with. These terms lead to three interesting effects. First, a spin-dependent tunneling term that couples to the electrons with a $\sigma_{x}$ ($\sigma_{y}$) matrix shifts the Weyl nodes on the $k_{x}$ ($k_{y}$) axis into opposite directions. Second, a spatial gradient in $\Delta_{x}$ ($\Delta_{y}$) as a function of the stacking direction $\hat{z}$, leads to a pseudo vector potential with a finite curl. The corresponding pseudo-field is $B_5\propto\pm\hat{y}$ ($B_5\propto\pm\hat{x}$).
Third, if the tunneling involves the Pauli-matrix $\sigma_{z}$, the Weyl nodes are tilted along the $k_z$-axis into opposite directions. The nodes can also be over-tilted into a type II regime if the spin-dependence of the tunneling is sufficiently strong, as seen in Fig .~\ref{fig:LC_Tilted}, which shows the band-structure of Eq.~\ref{eq:LCHamiltonain}.

We propose a system described by Eqs.~\ref{eq:LCHamiltonain} and \ref{eq:SpinDepTunTerms}, with tilted nodes due to a $\sigma_{z}$ spin-dependent tunneling across the trivial insulators, as a platform to test the predictions of the previous section. Note that in contrast to the situation discussed in section \ref{sec:ProbingOC}, where the nodes were tilted into the same direction, one now needs to apply a magnetic field to detect the chirality flip in the overtilted regime. Notice furthermore that the $\Delta_z$ term, in addition to controlling the tilt, also changes the energy of the Weyl nodes and the nodal separation. These changes in the spectrum may obscure the predicted differences in the optical conductivity measurement between the type I and II regimes. This issue can be resolved by fixing other parameters of the model as a function $\Delta_z$ in order to keep the energy of the nodes and their separation constant. For example, in a system with constant $\Delta_S$ and $\Delta_D$, a tilt of the nodes at constant nodal separation and energy at $E=0$ requires $m$ to be changed in correlation with the tilt $\Delta_z$ according to $m^2=\Delta_S ^2-\Delta_D ^2+\Delta_z^2$. The nodal location in this constrained case is given by 
\begin{eqnarray}
\begin{array}{c}
k_{0}=\pm\frac{1}{d}\left(cos^{-1}\left(-\frac{\Delta_{D}}{\Delta_{S}}\right)\right) ,
\end{array}
\label{eq:LCNodeLocationConstrained}
\end{eqnarray}
This kind of parameter engineering would allow to measure the optical conductivity while changing only the tilt of the nodes.

\subsection{Topological crystalline insulator multilayers}
The multi-layer construction discussed in the last section allowed for a system with the minimal number of two nodes because time reversal symmetry was explicitly broken by magnetization.
\begin{figure}[t!]
\captionsetup[subfloat]{labelformat=empty}
\captionsetup[subfloat]{farskip=-10pt,captionskip=-100pt}
\subfloat[]{\includegraphics[width=0.5\columnwidth]{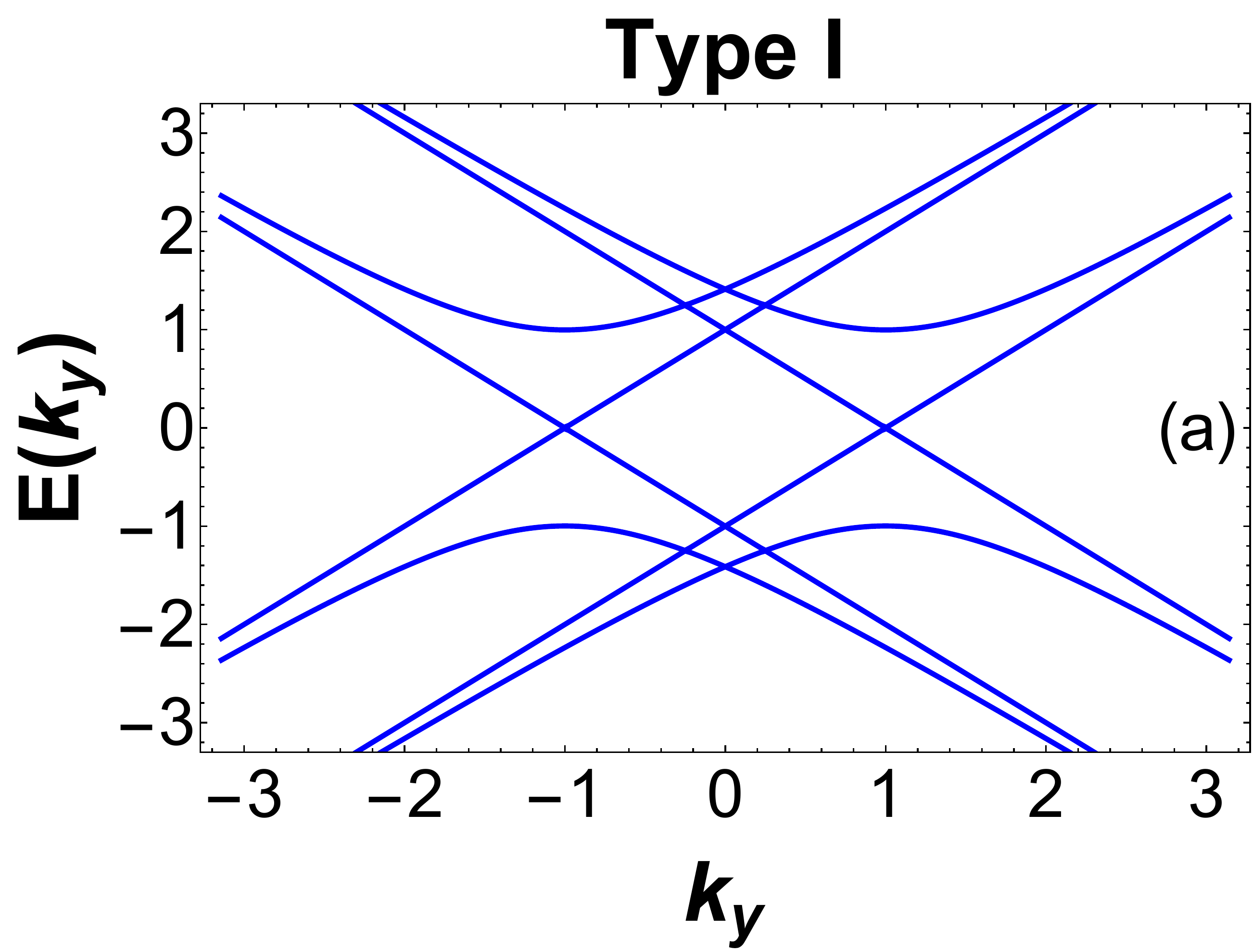}}
\subfloat[]{\includegraphics[width=0.5\columnwidth]{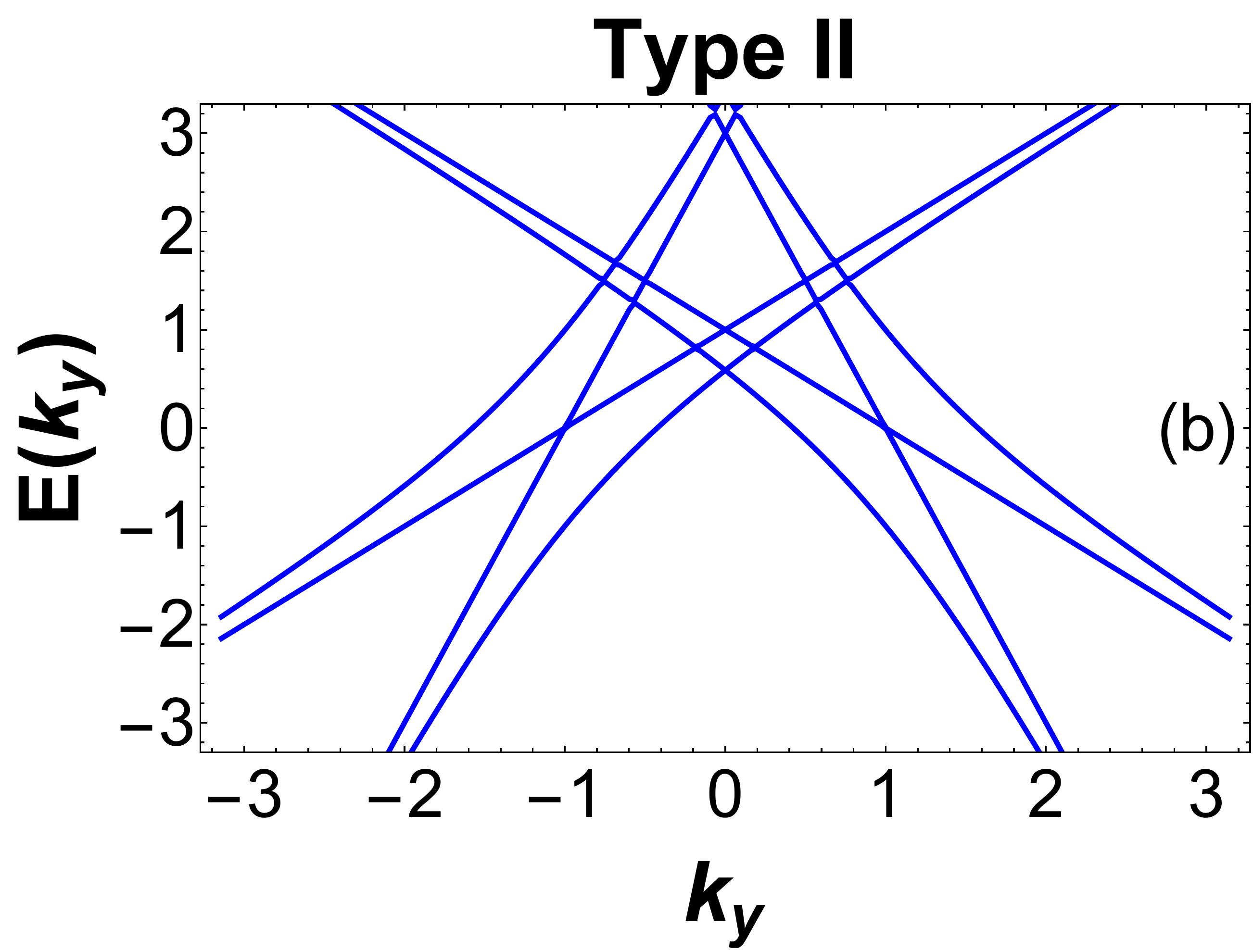}}
\vspace{-0.1in}\caption{\label{fig:TCI_MultiLayer} Band structure for the TCI-trivial insulator superlattice in Eq.~\ref{eq:TCIMultiLayerHamiltonainPeriodicBC}, for different values of TCI surface state tilt $g$. Parameters: $l=1$, $v_{F}=1$, $m=t_1=t_2=0.5$. The band structures are computed at $k_x=0$ and $k_z=2.1$. In the left panel (a) $g=0$, and in the right panel (b) $g=2v_{F}$. In the (b) the three-dimensional emerging cones are over tilted, a property inherited from the parent TCI. }
\end{figure}
As an alternative to that, we also propose a time-reversal symmetric system, where tilted cones are inherited from the parent material and need not be engineered by adjusting a hopping term. Following Ref.~\cite{Lau2017}, we propose a layered construction from alternating layers of topological crystalline insulators (TCIs) \cite{Fu2011} and ordinary insulators stacked in the $\hat{z}$ direction.
As it turns out, topological crystalline insulators can host over-tilted Dirac surface states \cite{Chiu2017,Pham2017,Wang2019}, predicted to exist in anti-pervoskite materials A$_3$EO, where A is an alkaline earth metal, and E denoted Pb or Sn \cite{Chiu2017}. Reference \cite{Chiu2017} discusses Ca$_3$PbO as a representative of this class of materials and shows that due to reflection symmetries, this material has non-trivial mirror Chern numbers on mirror planes, resulting in Dirac surface states with different tilts. In particular, the $(011)$ surface is predicted to host four Dirac cones, two of which are tilted in a direction that is parallel to a mirror symmetry line, thus preserving the symmetry.

We consider a time-reversal-invariant TCI protected by mirror symmetry $x\rightarrow-x$, where the mirror plane has Chern number $C_m=2$. Terminated in the $z$ direction, the surface of this TCI hosts two protected Dirac cones, which can be, for suitable material parameters,  over-tilted. The Dirac nodes in the $(k_x,k_y)$ plane of the surface reside at momenta $\pm\boldsymbol{d}=\left(0,\pm d\right)$. The tilt direction is parallel to the $yz$ mirror symmetry plane, preserving the symmetry. Using $s_{\mu}$, $\sigma_{\mu}$ and $\tau_{\mu}$ to represent Pauli matrices associated with the spin, top and bottom surface of the TCI and the two Dirac cones per surface degrees of freedom, respectively, and defining $\sigma_{\pm}=\left(\sigma_{x}\pm i\sigma_{y}\right)$ and $\tau_{\pm}=\left(\tau_{0}\pm \tau_{z}\right)$ (note that the definition of $\tau_{\pm}$ here is different then in the previous section), the Hamiltonian is given by

\begin{figure}[t]
\captionsetup[subfloat]{labelformat=empty}
\captionsetup[subfloat]{farskip=0pt,captionskip=1pt}
\subfloat[]{\includegraphics[width=0.8\columnwidth]{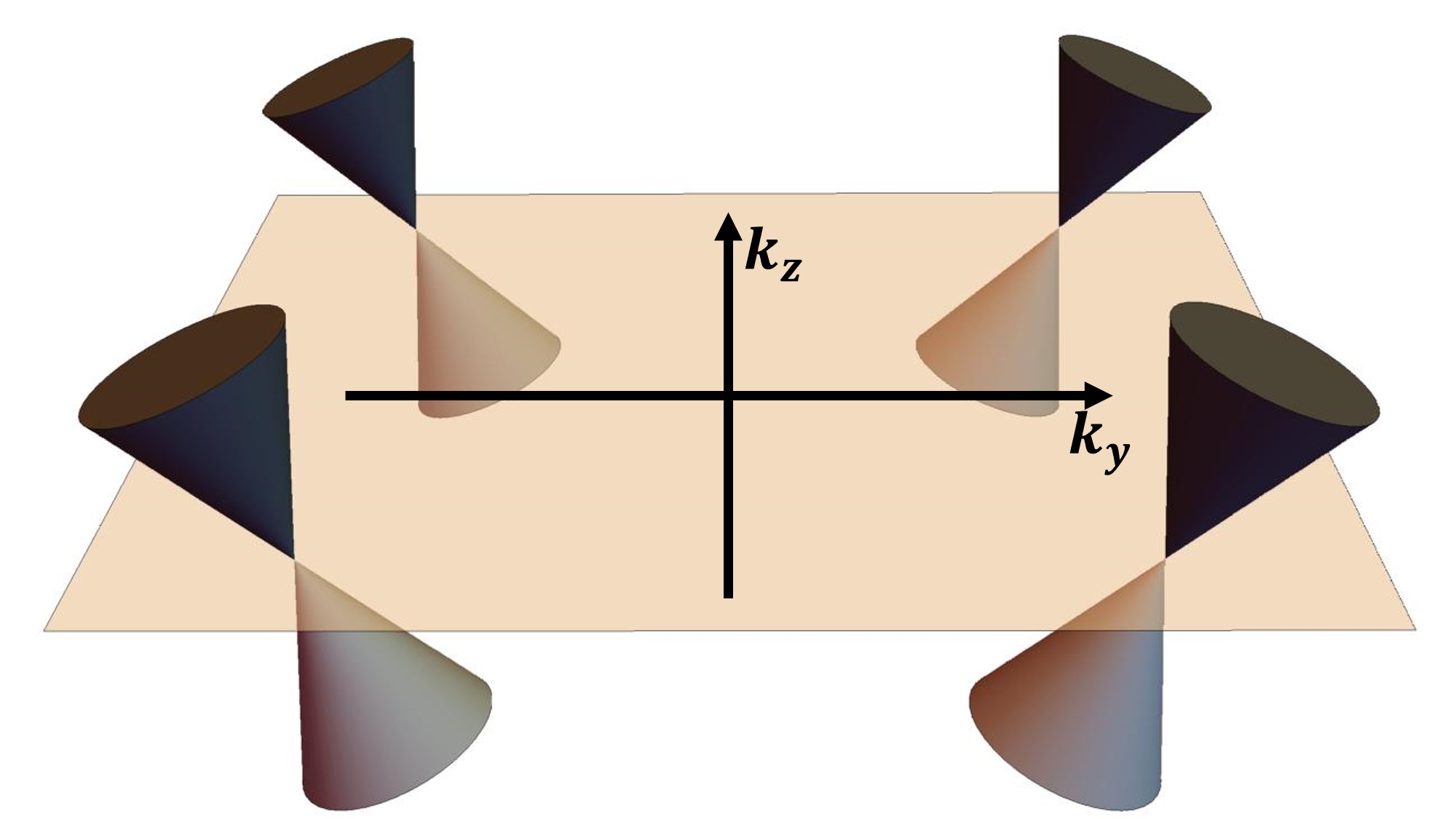}}
\vspace{-0.1in}\caption{\label{fig:TCI_Ilustration} Illustration of the arrangment of Weyl cones of the superlattice in Eq.~\ref{eq:TCIMultiLayerHamiltonainPeriodicBC}. The nodes are tilted along the $k_y$ axis. Creating a pseudo-field can be done by changing the location of the nodes via the mechanisms described in the main text, for example along the $z$ axis, as a function of $x$, leading to Landau levels dispersing along $k_y$.}
\end{figure}

\begin{widetext}
\begin{eqnarray}\nonumber
H(k)=
\frac{v_F}{2}\sigma_{z}\left(\hat{z}\times\boldsymbol{s}\right)\cdot\left[\tau_{+}\left(\boldsymbol{k_{\perp}}+\boldsymbol{d}\right)+\tau_{-}\left(\boldsymbol{k_{\perp}}-\boldsymbol{d}\right)\right]&+&
\frac{g}{2}\left[\tau_+\left(k_y+d\right)-\tau_-\left(k_y-d\right)\right]\\
&+&m\tau_{z}s_{z}+t_{1}\sigma_{x}+\frac{t_{2}}{2}\left(\sigma_{-}e^{ik_{z}l}+\sigma_{+}e^{-ik_{z}l}\right) ,
\label{eq:TCIMultiLayerHamiltonainPeriodicBC}
\end{eqnarray}
\end{widetext}
where the terms proportional to $v_F$ represent the Hamiltonian of a single TCI surface with un-tilted Dirac cones. The tilt of the surface states is accounted for by the terms $\propto g$. More precisely, these terms tilt the nodes in a direction parallel to the $\hat{k}_{y}$ axis and have an opposite sign for nodes at positive or negative $k_y$. This creates a tilt of the cones separated in $k_y$ into opposite directions with respect to each-other, preserving the $x\rightarrow-x$ reflection symmetry. For tilt amplitudes $|g|>v_F$ the cones are over-tilted. The mass $m$ opens a gap in the 2D Dirac surface states. This can be achieved, for example, by a mechanical strain or ferroelectric distortion, as suggested by Refs.~\cite{Lau2017,Hsieh2012}. Another path to opening a gap is by terminating the 3D TCI with a small angle with respect to the termination surface that supports the existence of the symmetry protected Dirac surface states. Finally, $t_{1}$ is the coupling amplitude between Dirac cones with the same momentum on the top and bottom surfaces of a single TCI layer, and $t_{2}$  is the coupling  amplitude for surfaces on adjacent layers. We assume periodic boundary conditions in the stacking direction $k_{z}$ and a layer width $l$. Depending on the microscopic parameters, this Hamiltonian can produce a Weyl semimetal with two pairs of over-tilted Weyl cones as shown in 
Fig. \ref{fig:TCI_MultiLayer}. For $m>0$ a finite range in parameter phase space emerges which supports a semimetallic phase~\cite{Lau2017}, which is unaltered by a tilt term. The arrangement of the cones in vicinity of the $(k_y,k_z)$ plane is illustrated in Fig.~\ref{fig:TCI_Ilustration}. 

This construction can be utilized to test the predictions of section \ref{sec:ProbingOC}, by applying a pseudo-field parallel to the tilt direction, i.e parallel to the $\hat{y}$ direction. Such a pseudo-field can be achieved by changing the $k_z$ separation of the Weyl nodes as a function of the $x$ coordinate, possibly by strain. This would lead to a reflection symmetry breaking, but considering that the $k_z$ separation is controlled by the gap opening term $m$, and that the gap can be tuned by the breaking of the TCI underlying reflection symmetry, this would lead to the desired effect. It is unclear how to control the magnitude of $m$ and in particular make it vary in a particular fashion in space. Therefore, we suggest to alternatively change the out of layer hopping element $t_2$, which also control the separation of the Weyl nodes in the $k_z$ direction, as a function of the $x$ coordinate. This can be done, for example, by changing the width of the trivial insulator layers as a function of $x$.

Given a pseudo-magnetic field along the $y$ direction, a pair of nodes which are separated in $k_z$ but share the same $k_x$ and $k_y$, will host a bulk zeroth pseudo-Landau level dispersing with the same sign along $k_y$. Due to the counter tilt of the TCI Dirac surface states, these pairs tilt oppositely (see Fig.~\ref{fig:TCI_Ilustration}). Naturally, time reversal symmetry imposes that the bulk zeroth Landau level from nodes representing time reversed partners disperse in opposite directions. This predicts the following results: when $|g|>v_F$, and depending on the direction of the pseudo-field, either \textit{all} four zeroth Landau levels flip together, or non of them do. As was explained in section \ref{sec:ProbingOC}, this leads to an asymmetry with respect to the pseudo-field direction in the optical conductivity measurement.

\section{Summary and Discussion}
To summarize, we have studied the effects of pseudo-magnetic fields on the chiral Landau levels of type II topological Weyl semimetals. As we have shown, for a simple situation of co-tilted cones in the presence of pseudo-fields, the optical conductivity can reveal the direction of the tilt due to chirality flips of the lowest Landau levels close to the nodes. In more complicated situations with multiple nodes, or when nodes are counter tilted, a clever combination of pseudo-fields and external fields can leave fingerprints of over-tilted cones in the optical conductivity. In addition, the realization that chirality flips of the lowest Landau levels, when not occurring in the bulk, could happen at the surface, may advance the detection of type II semimetals via surface sensitive probes such as ARPES. It is not, however, expected to have a significant contribution to the optical conductivity.  

We have presented two multilayer designs that can realize such semimetals and allow for the formation of pseudo-fields. In addition, it is possible for pseudo fields and tilted cones to emerge in twisted layered materials~\cite{Fengcheng2019}. In terms of actual materials, several layered materials such as MoTe$_2$ have the potential of realizing the physics we discuss here \cite{Sun2015,Deng2016,Jiang2017MoTe2}. Our work therefore addressed two outstanding challenges. To date, there is no clear signature of pseudo-fields in three dimensional materials, despite multiple works predicting their consequences \cite{Pikulin2016,Grushin2016,Liu2017,Gorbar2017,Gorbar2017b,Matsushita2018}. Evidence for their appearance has only been found in analogous topological metamaterials \cite{Peri2019,Jia2019}. In addition, distilling transport signatures of over tilted cones has not been achieved. We hope that this work could advance the field in both fronts.  

\section{Acknowledgments}
D.S. and R.I. are supported by the ISRAEL SCIENCE FOUNDATION (grant No. 1790/18). T.M. acknowledges financial support by the Deutsche Forschungsgemeinschaft via the Emmy Noether Programme ME4844/1- 1, the Collaborative Research Center SFB 1143, project A04, and the Cluster of Excellence on Complexity and Topology in Quantum Matter ct.qmat (EXC 2147). The authors would like to thank Emil Bergholtz for useful discussions.

\bibliography{draft}

\end{document}